\documentclass[12pt]{article}
\usepackage{a4wide}
\usepackage{amssymb}
\usepackage{graphicx}

\newcommand*{\R}{{\mathbb R}}
\newcommand*{\C}{{\mathbb C}}

\begin{document}
{\renewcommand{\thefootnote}{\fnsymbol{footnote}}
\begin{center}
{\LARGE  An algebraic approach to the ``frozen formalism''  problem of time}\\
\vspace{1.5em}
Martin Bojowald$^1$
and Artur Tsobanjan$^2$
\\
\vspace{0.5em}
$^1$ Institute for Gravitation and the Cosmos,
The Pennsylvania State
University,\\
104 Davey Lab, University Park, PA 16802, USA\\
\vspace{0.5em}
$^2$ King's College, 133 North River Street, Wilkes-Barre, PA 18711, USA\\
\vspace{1.5em}
\end{center}
}

\setcounter{footnote}{0}

\begin{abstract}
 The long-standing problem of time in canonical quantum gravity is the source of several
 conceptual and technical issues. Here, recent mathematical results are used
 to provide a consistent algebraic formulation of dynamical symplectic
 reduction that avoids difficult requirements such as the computation of a
 complete set of Dirac observables or the construction of a 
 physical Hilbert space. In addition, the new algebraic treatment makes it
 possible to implement a consistent realization of the gauge structure off the
 constraint surface. As a consequence, previously unrecognized consistency
 conditions are imposed on deparameterization---the method traditionally used to unfreeze
 evolution in completely constrained systems. A detailed discussion of how the
 new formulation extends previous semiclassical results shows that an internal time degree of freedom need
 not be semiclassical in order to define a consistent quantum evolution.
\end{abstract}

\section{Introduction}

General covariance gives rise to completely constrained systems whose canonical dynamics is
not driven by a Hamiltonian with respect to absolute time but rather by a
constraint function. The Hamiltonian constraint vanishes for all allowed sets of
phase-space variables, and it generates a canonical flow under which
observables are invariant. These requirements are the basis of several
conceptual and practical problems in the context of canonical quantum
gravity, including the problem of time.

Since observables are invariants of the flow, constructing them requires
integrating this flow, which, for a general dynamical system, is complicated
and can only be done locally. Even if constructed, such invariants pose an
interpretational difficulty: since they do not exhibit evolution in time, the
solution to the constrained system appears ``frozen''. As a way to address
both problems, observables can sometimes be interpreted as properties of
relational evolution. To this end, one phase-space variable is distinguished
as an internal time with respect to which other phase-space variables change
for given constant values of all observables \cite{GenHamDyn1,BergmannTime}. A
dynamical Hamiltonian picture can then be recovered, but in general only
locally because the internal time one picks may not be monotonically
increasing along the entire flow of the Hamiltonian constraint. In classical
mechanics, one can patch together finite pieces of overlapping internal-time
evolutions and obtain complete dynamical orbits.

Upon quantization, however, standard constructions pose several
difficulties, for instance because the usual requirement of unitary evolution
is in conflict with a time variable valid for just a finite range. Owing to
such problems, the general question of how to quantize completely constrained
systems has remained open, hampering constructions of canonical quantum
gravity in which this issue appears most prominently.

Questions remain even in cases in which a single internal time parameter may
suffice to describe the entire evolution. One problem that has not yet
received much attention is the fact that the usual procedures that implement
quantum constraints make it impossible to check whether off-shell properties
are correctly described.
There are mainly two different methods that can be used for an implementation
of constraints on a quantum theory, which in specific applications differ in
their tractability. For the discussion of off-shell properties and for later
reference, we briefly review the essential features of these two approaches.

Dirac quantization is one common approach to implementing
a quantum constraint. It starts with a kinematical Hilbert space on which the
basic phase-space variables of the system, such as positions and momenta of a
number of particles, are represented without considering the constraint at
this stage. The constraint $C$, given classically by a function of the basic
phase-space variables, is then turned into an operator $\hat{C}$ acting on the
same Hilbert space, where it is in general non-zero but has a non-trivial null
eigenspace of (generalized) states $|\psi\rangle$ that obey the condition
$\hat{C}|\psi\rangle=0$. If zero is in the discrete part of the spectrum of
$\hat{C}$, the corresponding eigenstates are normalizable and form a subspace
of the kinematical Hilbert space which can be identified with the physical
Hilbert space of the system on which the constraint is satisfied. Observables
acting on the physical Hilbert space need to keep this subspace invariant, and
are therefore required to commute with the constraint operator, a property
that defines Dirac observables. If zero is not in the discrete part of the
spectrum of $\hat{C}$, the corresponding eigenstates are not normalizable in
the kinematical Hilbert space. A separate physical Hilbert space must then be
constructed by introducing a suitable inner product on the space of
distributional solutions of $\hat{C}|\psi\rangle=0$ with a meaningful
representation of Dirac observables. In this case, the construction of a
physical Hilbert space is usually more ambiguous than in the case of zero in
the discrete spectrum of $\hat{C}$ and may require additional assumptions or
an application of different methods, such as those described in
\cite{Refined}.

The other frequently employed method, reduced phase-space quantization
\cite{ReducedDirac}, does not introduce a kinematical Hilbert space and avoids
the transition to a physical Hilbert space. Instead, it aims to construct the
physical Hilbert space directly by first solving the classical constraints
before quantization and then representing the resulting phase space on a
Hilbert space. The main difficulty is usually that the solution space to the
constraint may have a non-trivial topology, which makes quantization difficult
and can also introduce ambiguities. Moreover, it is necessary to parameterize
the solution space in a manner suitable for quantization, which in practice
requires a complete solution of all classical Dirac observables that have
vanishing Poisson brackets with the constraint.

While it remains unclear whether a reduced phase-space approach always agrees
with Dirac quantization, they do agree in many cases, with states and
operators of a physical Hilbert space corresponding to a quantization of the
classical constraint surface and its gauge flows. The end products of both of
these approaches describe the system confined to its classical constraint
surface and neither approach is therefore able to take into account behavior
of the constraint off that surface, which may be relevant in quantum systems
if the constraint surface is subject to fluctuations. (If zero is in the
discrete part of the spectrum of a constraint operator used for Dirac
quantization, the basic condition $\hat{C}|\psi\rangle=0$ implies that
fluctuations of the constraint vanish in any physical state. However, this
argument does not apply for zero in the continuous part of the spectrum.)  One
of the main observations of this paper is that off-shell properties are, in
fact, very relevant and can be used to place strong conditions on allowed
quantizations of constraints.

Testing off-shell properties is impossible on a Hilbert-space level that
strictly seperates kinematical and physical Hilbert spaces without
transformations between them. Moreover, the solutions of the constraint
equations used to define the physical Hilbert space are commonly obtained for
a single factor of a more complicated constraint $C$, writing $C=NC_H$ with
the flow rate $N$ (or lapse function in a gravitational context) and a reduced
constraint $C_H$. While $C_H=0$ implies $C=0$, depending on properties of $N$,
there may be subtle differences between the solution space and gauge
transformations of $C_H$ and $C$. In particular, if $N$ is not invertible, the
solution space of $C_H$ is smaller than the solution space of $C$. Moreover,
general gauge transformations, generated by products of operators $\Lambda_H C_H$ and
$\Lambda C=\Lambda N C_H$, respectively, behave differently because there may
be some $\Lambda_H$ that cannot be written in the form $\Lambda N$. Therefore,
while the solution space of $C_H$ is smaller, it is subject to more gauge
transformations. A single physical state with respect to $C_H$ then
corresponds to some non-trivial subset of physical states with respect to $C$,
and not all physical states of $C$ can be related to physical states of
$C_H$. While phase-space regions where $N$ is not invertible may easily be
treated as special cases in a classical procedure, the quantum behavior is
more complicated.

In~\cite{AlgebraicTime}, we introduced a new algebraic approach to
quantizing systems with a Hamiltonian constraint in which we avoid the thorny issues around the physical Hilbert space, by postponing its construction for as long as possible. At
the same time, in order to facilitate comparisons between different choices of
internal times and their corresponding deparameterizations, we aimed to
formulate all relevant structures related to the constraint surface, gauge
transformations, observables, and evolution on a single mathematical objects
and naturally derived features. Using algebraic methods, all constructions are
based on the original algebra $\mathcal{A}$ of kinematical observables and a
constraint element $C\in\mathcal{A}$, as well as specific ideals of
$\mathcal{A}$, factor algebras, and homomorphisms between them. A number of
independent and largely uncontrolled choices that are required in the
traditional construction of a physical Hilbert space, mainly its inner product
if the constraint operator has zero in its continuous spectrum, can then be
avoided.

The new treatment revealed several new properties of quantum relational
evolution that are important for physical applications. In particular,
Hamiltonian constraints typically encountered in models of quantum gravity,
which in general are quadratic in momenta, are subject to previously
unrecognized restrictions on their factor ordering for relational evolution to
exist at a mathematically rigorous level. They are particularly strong in
cases of constraints in which the term quadratic in momenta is multiplied by a
phase-space dependent lapse function, as is common in gravitational systems
where metric components appear in kinetic energies. Our new restrictions may
help to reduce quantization ambiguities, but in some cases they may also
eliminate relational evolution altogether, at least in the strict algebraic
form. In this way, our mathematical discussion serves to highlight important
choices that must be made in quantum symplectic reduction and the ambiguities that they introduce,
placing more control on the traditional treatment in which it is difficult,
for instance, to parameterize the freedom involved in choosing an inner
product for the physical Hilbert space.

The rest of our presentation is structured as
follows. Section~\ref{sec:AlgebraicApproach} reviews the algebraic perspective
on constrained quantum systems and their reduction based on the construction
of Dirac observables that was introduced
in~\cite{AlgebraicTime}. Section~\ref{sec:GaugeSections} discusses the
alternative approach to characterizing the gauge freedom that remains after a
constraint is solved algebraically, assuming that, as is often the case, Dirac
observables are not available, culminating in an algebraic definition of
deparameterization. Section~\ref{sec:Factorization} treats constraints that
cannot themselves be deparameterized relative to a given clock, but possess
deparameterizable factors. We find that only a very restricted class of
constraints can be deparameterized either directly or by factorization. Viable
approximate methods for deparameterizing other constraints are discussed in
section~\ref{sec:Approximations}. Section~\ref{sec:HSpace} briefly addresses
the link between algebraic states that we employ throughout our construction
and states in a Hilbert space representation of a constrained
system. Section~\ref{sec:Implications} explores the implications of our
results.

Sections~\ref{sec:AlgebraicApproach} to \ref{sec:Factorization} review
material published elsewhere, mainly in \cite{AlgebraicTime}, but in a manner
that is more accessible to a physics audience. In particular, we focus on
essential and conceptual features rather than detailed assumptions necessary
for rigorous proofs, and we present a streamlined result that is not as
general as those of \cite{AlgebraicTime} but serves to highlight new
properties. The final sections, \ref{sec:Approximations} to
\ref{sec:Implications}, contain entirely new material.

\section{Algebraic treatment of a single quantum constraint}
\label{sec:AlgebraicApproach}

For our purposes the (kinematical) degrees of freedom of a quantum system are
described by an associative, complex, unital $*$--algebra, which we will
denote by $\mathcal{A}$. The $*$-operation, mapping any element
$A\in\mathcal{A}$ to another element $A^*\in\mathcal{A}$ such that $A^{**}=A$
(as well as $(A+B)^*=A^*+B^*$ and $(AB)^*=B^*A^*$) defines an analog of
Hermitian conjugation at the algebraic level. As usual, true physical
observables of the system correspond to $*$--invariant elements of
$\mathcal{A}$.

\subsection{States}

Measurement results of observables are given by numbers rather than algebra
elements. For physical interpretations, it is therefore necessary to introduce
suitable mappings from the algebra to, in general, complex numbers. The latter
are interpreted as expectation values of the observable in a state defined by
the mapping $\omega$ from $\mathcal{A}$, $\omega(A)=\langle A\rangle$. The
mapping should therefore be linear. Moreover, physical expectation values of
various operators in a given state are not arbitrary but restricted by
uncertainty relations. These relations, in their usual derivation, follow from
Cauchy--Schwarz inequalities, which in turn are implied by a positivity
condition on states $\omega$: A linear functional
$\omega : \mathcal {A} \rightarrow \mathbb{C}$\ is positive if
\[
\omega \left(AA^*\right) \geq 0 \quad \mbox{for all}\quad A\in \mathcal{A} \ .
\]
The positivity condition implies the desired Cauchy--Schwarz inequality
\begin{equation}\label{CS}
  |\omega(AB^*)|^2 \leq |\omega(AA^*)| |\omega(BB^*)|
\end{equation}
as well as, for a unital algebra as assumed here,
\[
  \omega(A) = \overline{\omega(A^*)} \quad\mbox{for all}\quad A, \, B \in
  \mathcal{A}
\]
using the complex conjugate $\bar{a}$ of $a$. In particular, expectation
values of $*$-invariant $A$ are real.
If the algebra is represented on a Hilbert space, a state is commonly given by
an element of the Hilbert space up to normalization (a pure state or wave
function $\psi$), or by a density operator acting on the Hilbert space (a
mixed state or density matrix $\hat{\rho}$). Both examples obey the conditions
for an algebraic state. In general, an algebraic state may therefore be mixed,
but it is defined even if there is no representation on a Hilbert space and is
therefore more general.

Positive linear functionals describe possible outcomes of physical
measurements that are usually expressed through the construction of a Hilbert
space on which wave functions or density matrices are defined to represent
states. The concept of positive linear functionals on an algebra is, in fact,
closely related to the concept of Hilbert-space representations because,
according to the Gelfand-Naimark-Segal theorem, every such representation of
$\mathcal{A}$\ can be constructed by starting with an appropriate positive
linear functional on $\mathcal{A}$,
at least in the case of a $C^*$-algebra (which has a suitable norm): Every
algebra is also a vector space, which can be used as the vector space
underlying a Hilbert space. An algebraic state may be used to introduce an
inner product on a suitable factor space of the original vector space by first
constructing the sesquilinear form $\langle A,B\rangle=\omega(A^*B)$ for
$A,B\in\mathcal{A}$. By positivity of the state, the sesquilinear form is
semi-definite and therefore defines a unique inner product on the factor space
in which we factor out zero-norm states, given by all $A\in\mathcal{A}$ such
that $\omega(A^*A)=0$ (defining a left ideal in the algebra). A Hilbert space
is obtained by completion of the factor space. Since pure states in the
Hilbert space are given by algebra elements modulo the ideal, multiplication
in the algebra defines a Hilbert-space representation of the algebra. If this
representation is irreducible, the state $\omega$ is pure. See for instance
\cite{LocalQuant} for a discussion.

In a sense, therefore, the space of all possible positive
linear functionals on $\mathcal{A}$\ contains all possible representations of
the quantum system.
However, in the context of constrained quantization with its distinction
between kinematical and physical Hilbert spaces, the positivity condition may
take different forms depending on whether the constraint has been imposed
yet: Moving to a physical Hilbert space that is not a subspace of the
kinematical Hilbert space means that there is no obvious and unambiguous
relationship between the two inner products. Since the physical inner product
is relevant for observations while the kinematical inner product is rather an
intermediate construct on the way to the physical Hilbert space, we will drop
the positivity condition on kinematical states and implemented only when we
are at a stage comparable to the physical representation. Accordingly,
we will use $\Gamma$\ to denote the space of all complex linear functionals on
$\mathcal{A}$\ that are normalized, that is $\omega(\mathbf{1}) = 1$, and we
will refer to elements of $\Gamma$\ as states even if they are not
positive. Note that, with the normalization condition, $\Gamma$\ is not a
vector subspace of the space of all linear functionals, however it is closed
with respect to normalized sums
$(a_1 \omega_1 + a_2 \omega_2 + \ldots + a_N \omega_N) \in \Gamma$, as long as
$(a_1+a_2 + \ldots a_N) = 1$.  Physical states will belong to some Hermitian
representation of $\mathcal{A}$\ and will therefore be positive.

In this treatment, unless explicitly stated otherwise, we will use the
so-called Schr\"odinger picture of time evolution, where states evolve with
time, while operators that are not explicitly time-dependent remain fixed. The
most common way to specify time dependence of a quantum system is through the
commutator with a Hamiltonian operator
\begin{equation} \label{eq:Hdynamics}
\frac{{\rm d}}{{\rm d}t} \omega_t(B) = \frac{1}{i\hbar} \omega_t \left( [B, H] \right)
\end{equation}
for $B\in\mathcal{A}$.
We treat the above relation as a differential equation to be solved for the
one-parameter family of states $\omega_t$, $t\in \mathbb{R}$. From the
algebraic perspective, (\ref{eq:Hdynamics}) is a prescription for constructing
an infinite system of coupled ordinary differential equations, since, in order
to find $\omega_t(B)$\ we also need $\omega_t([B, H])$,
$\omega_t([[B, H], H])$, etc. We will, in general, not attempt to integrate
such flows explicitly, however, under the assumption that this system
possesses a unique solution for a given $\omega_0$, purely algebraic methods
can be used to deduce interesting properties of the integrated flow. For
example, Lemmas 1 and 2 in~\cite{AlgebraicTime} show that, provided $\omega_t$\
is positive and normalized for some $t_0$, this property is preserved along
the entire dynamical flow.

\subsection{Physical states on a quantum system with a single constraint}\label{sec:PhysicalStates}

We assume that the unconstrained system has a well-defined quantization that
results in an associative, complex, unital $*$--algebra $\mathcal{A}$, which
we will call the kinematical algebra. The system is subject to a single
constraint, represented by a distinguished kinematical element
$C \in \mathcal{A}$, such that $C=C^*$, $C$\ does not possess an inverse, and
is not a divisor of zero within $\mathcal{A}$, so that $AC=0$\ implies
$A=0$. (The system may also possess a Hamiltonian distinct form $C$, as we are
not yet specifically considering the case of a completely constrained system.) 
We begin with several definitions. In line with our earlier
discussion, the {\bf space of kinematical
  states}, denoted $\Gamma$, is the space of all complex-linear functionals
on $\mathcal{A}$, which are normalized, $\omega(\mathbf{1})=1$, but not, in
general positive. A state $\omega \in \Gamma$ is called a {\bf solution of the constraint} if: $\omega(AC) = 0$ for all
$A \in \mathcal{A}$. The {\bf constraint surface} $\Gamma_C\subset \Gamma$ is
the space of all solutions of $C$.

Any element $A\in \mathcal{A}$\ generates a flow $S_A(\lambda)$ on $\Gamma$ analogous to
(\ref{eq:Hdynamics}) but with $H$ replaced by $A$:
\begin{equation}\label{eq:Flow}
  i\hbar \frac{{\rm d}}{{\rm d} \lambda} \left( S_A(\lambda) \omega(B) \right) :=
  S_A(\lambda) \omega([B, A]) \\ , \ \  {\rm and}\ \ S_A(0) = {\rm id}\ .
\end{equation}
Since $A$ is not in general a Hamiltonian, this flow is not in general temporal. In fact, if we set $A=C$\ the corresponding flow is the gauge flow of the constraint, which keeps physical properties unchanged. States related by gauge flows should therefore be indistinguishable by measurements. In general, for any $A\in \mathcal{A}$ the product $AC$\ should generate a gauge flow, because a state that solves the constraint $C$ also solves the constraint $AC$ (in this ordering). If
$A\not=0$ and $A\not=\mathbf{1}$, the flow of $AC$ is in general non-trivial
and independent of the flow of $C$ (using the assumption that $C$ not be a
divisor of zero). This property is mathematically expressed by an equivalence
relation: A pair of states $\psi, \omega \in \Gamma$\ are said to be {\bf
  $C$--equivalent}, $\omega \thicksim_C \psi$, if there exist
$A_1, A_2, \ldots, A_N \in \mathcal{A}$, as well as
$\lambda_1, \lambda_2, \ldots \lambda_N \in \mathbb{R}$, such that
\[
\psi = S_{A_1C} (\lambda_1) S_{A_2C} (\lambda_2) \ldots S_{A_NC} (\lambda_N) \omega \ .
\]
We denote the entire orbit generated by all of the constraint flows from some
state $\omega \in \Gamma$\ as
$[\omega]_C := \{ \psi \in \Gamma \colon \psi \thicksim_C \omega \}$. The
constraint surface $\Gamma_C$\ is preserved by the flows induced by all
constraint elements $AC$ (Lemma 4 in~\cite{AlgebraicTime}), so that for any
$\omega \in \Gamma_C$\ the orbit is entirely contained within the constraint
surface $[\omega]_C \subset \Gamma_C$.

Observables of the constrained system are, by
definition, invariant under any gauge flow, (\ref{eq:Flow}) implies that
observables must commute with $C$: The {\bf observable algebra}
$\mathcal{A}_{\rm obs}$ is the commutant of $C$,
$\mathcal{A}_{\rm obs}:=\{A\in\mathcal{A} \, :\, [A, C] = 0 \}$.
It follows that $\mathcal{A}_{\rm obs}$\ is a unital $*$--subalgebra of
$\mathcal{A}$ (Lemma 3 in~\cite{AlgebraicTime}). Moreover, a pair of $C$--equivalent
states in $\Gamma_C$\ assign identical values to the elements of
$\mathcal{A}_{\rm obs}$ (Lemma 5 in~\cite{AlgebraicTime}). These results
motivate the following definition: The {\bf physical space of states}
$\Gamma_{\rm phys}$, is the space of $C$--equivalence classes of states on
$\mathcal{A}$, which solve the constraint. In other words
$\Gamma_{\rm phys} =\Gamma_C/\thicksim_C$. Just like $\Gamma$\ and $\Gamma_C$,
$\Gamma_{\rm phys}$\ is closed with respect to normalized sums.

In theory, $\mathcal{A}_{\rm obs}$\ and $\Gamma_{\rm phys}$\ together comprise
the algebraic solution to the quantum constraint, and we can naturally
restrict physical states to the ones that are positive on
$\mathcal{A}_{\rm obs}$\ with respect to its inherited
$*$--structure. Moreover, if the kinematical algebra possesses a distinguished
Hamiltonian element $H\in \mathcal{A}$\ such that $[H, C] = 0$, then
$H \in \mathcal{A}_{\rm obs}$, and it can be used to generate the dynamical
flow on $\Gamma_{\rm phys}$\ via the commutator as in
equation~(\ref{eq:Hdynamics}).\footnote{If $[H, C] \neq 0$, then, in order for
  $H$\ to generate dynamics that preserve $\Gamma_{\rm phys}$, we need to
  impose additional constraints $[H, C]$, $[H, [H, C]]$\ and so on. These
  constraints are independent of $C$ unless $\omega([H,C])=0$ for all
  $\omega\in\Gamma_C$. Systems with multiple constraint elements are outside
  of the scope of the present manuscript and will be discussed elsewhere.}

\subsection{A simple example of an algebraic constraint}\label{sec:NonHamCExample} 

As an example we consider a quantum particle kinematically free to move in
two-dimensions but restricted to one dimension by a constraint. For
simplicity, we also pick a rather artificial kinematical Hamiltonian element,
which consists of a harmonic potential and a kinetic energy that has no
dependence on the momentum component in the restricted direction. Let the
kinematical algebra $\mathcal{A}$\ consist of all complex polynomials in the
basic elements $Q_1$, $P_1$, $Q_2$, $P_2$, and $\mathbf{1}$, where the
generating elements are star-invariant, $Q_1=Q_1^*$\ etc., and are subject to
the usual canonical commutation relations (CCRs), where the only non-trivial
commutators are $[Q_1, P_1] = [Q_2, P_2] = i \hbar \mathbf{1}$\,. Let the
Hamiltonian element be
\[
  H = \frac{1}{2m} P_1^2 + \frac{k}{2} \left( Q_1^2 + Q_2^2 \right)
\]
(where $k$\ and $m$\ are some positive constants with suitable units), and let
our particle be subject to the constraint $C=Q_2$, which classically restricts
its motion to $Q_2=0$ and eliminates $P_2$ if the gauge flow us factored out.

Due to the form of the constraint, it is convenient to write elements of
$\mathcal{A}$\ as linear combinations of \emph{specially-ordered} monomials
$P_1^{n_1}P_2^{n_2} Q_1^{l_1}Q_2^{l_2}$ (with
$P_1^{0}P_2^{0} Q_1^{0}Q_2^{0} = \mathbf{1}$), so that any $A\in \mathcal{A}$\
can be written as a finite sum
\[
A = \sum_{n_1,\, n_2,\, l_1,\, l_2 = 0} a_{l_1l_2}^{n_1n_2}\, P_1^{n_1}P_2^{n_2} Q_1^{l_1}Q_2^{l_2} \ ,
\]
for some $a_{l_1l_2}^{n_1n_2} \in \mathbb{C}$, where $n_i$\ and $l_i$\ terminate at finite maximum values. This linear decomposition is unique for each element
$A \in \mathcal{A}$\ because the set of specially-ordered monomials
$\{P_1^{n_1}P_2^{n_2} Q_1^{l_1}Q_2^{l_2}\}$\ is linearly independent. Due to
the form of the CCRs, $C$\ commutes precisely with the specially-ordered
monomials for which $n_2=0$, therefore the observable algebra here consists of
linear combinations of $P_1^{n_1} Q_1^{l_1}Q_2^{l_2}$. Note that $[H, C] = 0$\
and therefore $H \in \mathcal{A}_{\rm obs}$.

Linear states on $\mathcal{A}$\ are completely characterized by the values
they assign to the linear basis $\{P_1^{n_1}P_2^{n_2} Q_1^{l_1}Q_2^{l_2}\}$:
\[
  \omega(A) = \sum_{n_1,\, n_2,\, l_1,\, l_2 = 0}
  a_{l_1l_2}^{n_1n_2}\, \omega \left( P_1^{n_1}P_2^{n_2} Q_1^{l_1}Q_2^{l_2}
  \right) \ .
\]
Any $\omega \in \Gamma$\ needs to be normalized $\omega (\mathbf{1} ) = 1$. If
we wanted to restrict to kinematically positive states we would have to
enforce an infinite set of additional conditions on the values assigned to
these basis elements, such as $\omega(Q_1) \in \mathbb{R}$,
$\omega\left( Q_1^2 \right) \geq 0$, and generalizations of uncertainty
relations
\[
  \left( \omega\left( Q_1^2 \right)- \omega\left( Q_1 \right)^2\right) \left(
  \omega\left( P_1^2 \right)- \omega\left( P_1 \right)^2\right) -(
\omega\left( P_1 Q_1 \right)-\omega ( P_1)\omega ( Q_1)+i\hbar/2)^2 \geq
\hbar^2/4\,.
\]
However, in this treatment, we do not impose positivity on kinematical states;
moreover, there are choices of ordering that are more convenient for imposing
positivity than the one selected above.

Using the specially-ordered basis, $\Gamma_C$, the set of solutions to the
constraint, consists of the states that satisfy $\omega_{l_1l_2}^{n_1n_2}=0$\
for all $l_1, l_2, n_1$ whenever $l_2\neq 0$. The flows generated by the
constraint element through the commutator may be characterized by the way
in which they affect the values a state assigns to each basis monomial,
\begin{eqnarray*}
  \frac{{\rm d}}{{\rm d} \lambda} \left. \left( S_{AC} (\lambda) \omega \right)\left(
  P_1^{n_1}P_2^{n_2} Q_1^{l_1}Q_2^{l_2} \right) \right|_{\lambda = 0} &=&
                                                                          \omega \left( \left[ P_1^{n_1}P_2^{n_2} Q_1^{l_1}Q_2^{l_2}, AC \right] \right) 
  \\
                                                                      &=&
                                                                          \omega \left( \left[ P_1^{n_1}P_2^{n_2} Q_1^{l_1}Q_2^{l_2}, A \right] Q_2 \right) \\ &&- i\hbar n_2 \, \omega \left( A P_1^{n_1}P_2^{n_2-1} Q_1^{l_1}Q_2^{l_2} \right) \ . 
\end{eqnarray*}
For any $\omega \in \Gamma_C$, the first term in the last expression
identically vanishes, while the second term is only non-zero if both
$n_2 \neq 0$\ and $l_2=0$. Thus, the values $\omega \in \Gamma_C$\ assigns to
elements of $\mathcal{A}_{\rm obs}$ (that is
$\omega\left( P_1^{n_1} Q_1^{l_1}Q_2^{l_2}\right)$) are unaffected by the
constraint-induced flows (see Lemma~5 in~\cite{AlgebraicTime}). Therefore,
solution states that are distinct when restricted to $\mathcal{A}_{\rm obs}$\
correspond to distinct elements of $\Gamma_{\rm phys} = \Gamma_C/\thicksim_C$.

Since each physical state annihilates elements of $\mathcal{A}_{\rm obs}$\
that have the form $AC \equiv AQ_2$, it is more accurate to say that they are
states on the quotient $\mathcal{A}_{\rm obs}/\mathcal{A}_{\rm obs}C$, rather
than the full observable algebra. Here, it is straightforward to explicitly
verify that $\mathcal{A}_{\rm obs}C$\ is a two-sided $*$--ideal of
$\mathcal{A}_{\rm obs}$, which naturally makes the quotient
$\mathcal{A}_{\rm obs}/\mathcal{A}_{\rm obs}C$\ into a $*$--algebra isomorphic
to the algebra $\mathcal{B}$\ of polynomials in $Q_1$, $P_1$, and
$\mathbf{1}$\ only, under the mapping
\[
\eta\colon \left[ P_1^{n_1} Q_1^{l_1}Q_2^{l_2} \right] \mapsto P_1^{n_1} Q_1^{l_1} \delta_{0\, l_2} \ ,
\]
extended to the entirety of $\mathcal{A}_{\rm obs}/\mathcal{A}_{\rm obs}C$\ by
linearity. Here $[A]$\ denotes the coset of $A \in \mathcal{A}_{\rm obs}$\
with respect to the ideal $\mathcal{A}_{\rm obs}C$. Verifying that $\eta$\ is
a $*$--algebra isomorphism is straightforward. This mapping also identifies a
Hamiltonian $\eta(H) = \frac{1}{2m} P_1^2 + \frac{k}{2} Q_1^2 $, which
generates time-evolution on
$\mathcal{B} \cong \mathcal{A}_{\rm obs}/\mathcal{A}_{\rm obs}C$\ through the
commutator.

To summarize, we used the \emph{specially-ordered} linear basis on the
kinematical algebra in order to construct both the observable algebra and the
space of constraint solutions. Since $C$--equivalent solution states assign
identical values to all elements of $\mathcal{A}_{\rm obs}$, physical states
can be distinguished by the values they assign to $\mathcal{A}_{\rm obs}$. On
the other hand, two elements of $\mathcal{A}_{\rm obs}$\ that differ by an
element of $\mathcal{A}_{\rm obs}C$\ will be assigned identical values by all
physical states. We therefore characterize physical states by the values they
assign to $\mathcal{B} \cong \mathcal{A}_{\rm obs}/\mathcal{A}_{\rm obs}C$,
which, in this simple case, comes equipped with a physical Hamiltonian. At
this point, the construction of physical Hilbert space and physical dynamics
can proceed directly from $\mathcal{B}$\ by starting with a suitable positive
state.

\subsection{Important limitations of using the observable algebra}\label{sec:AobsDifficulties}

In general, it may not be feasible to characterize the physical states of a
constrained system by first identifying the corresponding observable
algebra. The artificial simplicity of the explicit example from the previous
section has allowed us to temporarily sweep several important difficulties
under the rug; we list them below.
\begin{enumerate}
\item Perhaps most obviously, the simplicity of $\mathcal{A}$\ and $C$\ has
  allowed us to infer the observable algebra explicitly more-or-less ``by
  inspection''. As far as we know, no universal method for constructing the
  commutant of $C$\ within an arbitrary $*$--algebra exists.
\item A more subtle caveat is that, even in this simple example, it is not
  obvious that the observable algebra $\mathcal{A}$, defined as the subalgebra
  of those $A$ that commute with $C$, necessarily resolves the physical
  states, defined as $C$-equivalence classes of states on $\mathcal{A}$. While
  it is straightforward to see that any physical state corresponds to a unique
  state on $\mathcal{B}$ in the preceding example, the converse is not
  necessarily true. Since we have not characterized the $C$--equivalence
  classes on $\Gamma_C$\ here, we cannot ascertain that it is possible to
  distinguish any two distinct physical states through the values they assign
  to $\mathcal{B}$.
\item There is another way in which $\mathcal{A}_{\rm obs}$ may end up being too small to be able to resolve all physical states. The invariants with respect to the adjoint
  constraint action, $[\cdot,C]$, may inhabit an enlargement of $\mathcal{A}$,
  such as infinite power series in elements of $\mathcal{A}$\ that converge in
  a suitably-defined sense. In practice, extensions of the original algebra would be constructed on suitable Hilbert-space representations. However, such definitions of the observable algebra generally depend on
the chosen representation and introduce additional quantization ambiguities.
\item Even if one allows extensions of the kinematical algebra, a sufficient
  number of invariants may not exist at all, as would be expected if the
  classical flow of $C$\ is non-integrable
  \cite{DiracChaos,DiracChaos2}.\footnote{For example, the two-dimensional
    kinematical system from the previous section could be subjected to
    $C=\frac{1}{2}P_1^2 + \frac{1}{2}P_2^2 + \frac{1}{2}Q_1^2 +
    \frac{1}{2}Q_2^2 + Q_1^2Q_2-\frac{1}{3}Q_2^3$, which comes from the
    H\'enon-Heiles Hamiltonian and generates a classically non-integrable
    flow.}
\item Even with a single primary constraint, carrying the dynamics over to the
  physical space can get complicated if $[H, C] \neq 0$\ and secondary
  constraints need to be imposed. This will necessitate a sharper definition
  of $\mathcal{A}_{\rm obs}$\ applicable within this context.
\end{enumerate}

\section{Algebraic gauge fixing}\label{sec:GaugeSections}

The approach developed in~\cite{AlgebraicTime} avoids some of the difficulties
associated with constructing physical observables by focusing on
characterizing the physical states instead. We note that, aside from the
requirement that physical states are positive on $\mathcal{A}_{\rm obs}$,
$\Gamma_{\rm phys}$\ can be constructed quite independently of
$\mathcal{A}_{\rm obs}$. Schematically, we first pass to the quantum
constraint surface $\Gamma_C$, which imposes a set of algebraic conditions on
the values assigned by the states. The $C$--equivalence relation generates
orbits on $\Gamma_C$, with each distinct orbit corresponding to a distinct
physical state. Since all points on a given orbit correspond to the same
physical state, we refer to the freedom to move along an orbit as \emph{gauge}
freedom. This procedure makes no use of the properties of
$\mathcal{A}_{\rm obs}$. In the course of our developments, we will see that
even the positivity condition can be spelled out independently of
$\mathcal{A}_{\rm obs}$.

We note in passing that the algebraic method for fixing quantum gauge freedom based on an internal clock described in the rest of this section appears to be related to the Hilbert space and operator construction of temporal quantum reference frames in~\cite{QuantumRefSwitch,QuantumRef4} though the full details of this relation are yet to be understood.

\subsection{A geometrical picture}
\label{sec:Geometry}

In classical theories with gauge freedom the state space is a symplectic or
Poisson manifold with gauge orbits forming lower-dimensional embedded
surfaces. Classical gauge freedom can be completely fixed by specifying a surface
that intersects each gauge orbit at one point---the main idea of fixing the
gauge is to separate orbits by these points of intersection. Points on the
gauge-fixing surface are, of course, by inclusion, also points of the original
state space, corresponding to a given set of values of the gauge-fixing
functions. These points can be distinguished from each other by the values
they assign to functions that ``live'' \emph{on} the gauge-fixing
surface. Intersections between gauge orbits and a (possibly partial)
gauge--fixing surface provide a representative subset of states from each
orbit.

By direct analogy, as a form of ``gauge-fixing'' we could, in some
yet-to-be-determined way, select a subset of states $\Theta \subset \Gamma_C$\
to represent the $C$--equivalence classes of states on $\Gamma_C$\ and attempt
to implement one or more of the following highly desirable properties.
\begin{enumerate}
\item If the gauge freedom is completely fixed, then each element of
  $\Gamma_{\rm phys}$\ that has a representative in $\Theta$\ will only have
  one such representative. That is, for any $\omega \in \Gamma_C$, we desire
  that $\Theta \cap [\omega]_C$\ contains no more than one element.
\item The selected subset $\Theta$\ should have at least one representative
  from each physical state. In other words, we desire that
  $\Theta \cap [\omega]_C$\ contains at least one element for each
  $\omega \in \Gamma_C$.
\item \label{lst:gauge} Finally, if the gauge-fixed theory is to be physically
  interpretable, the collection of states in $\Theta$\ should be related to
  positive states on some unital $*$--algebra (replacing ${\cal A}_{\rm obs}$)
  that, in turn, holds some relation to the original kinematical degrees of
  freedom of the system studied.
\end{enumerate}

There is an immediate difficulty with item~(\ref{lst:gauge}) above. The states
within $\Gamma_C$\ itself are most naturally identified as states on the
linear quotient space $\mathcal{A}/\mathcal{A}C$: Since such states, by
definition, annihilate any algebra element of the form $AC$ with some
$A\in\mathcal{A}$, they are uniquely defined on the equivalence classes
$[B]=B+\mathcal{A}C$ that define the factor space
$\mathcal{A}/\mathcal{A}C$. However, while $\mathcal{A}C$\ is a subalgebra of
$\mathcal{A}$, it is only a left-sided ideal because
$(AC)B \notin \mathcal{A}C$ in general. In addition, its is not guaranteed to
be $*$--invariant because $(AC)^*=CA^* \notin \mathcal{A}C$ in
general. Therefore $\mathcal{A}/\mathcal{A}C$\ inherits neither the full
multiplicative structure, nor the $*$--structure from $\mathcal{A}$. This means that there is no obvious way to interpret $\Gamma_C$\ as the complete collection of states of some (reduced) system or to impose positivity via a $*$--operation. We can
try to remedy this situation by first identifying elements of $\Gamma_C$\ with
states on some suitably chosen $*$--algebra, before attempting to fix any
gauge freedom. The usual treatment of quantum constrained systems in terms of
the observable algebra $\mathcal{A}_{\rm obs}$ has no analog of this
intermediate stepping stone.

\begin{figure}[h]
\begin{center}
  \includegraphics[scale=.65]{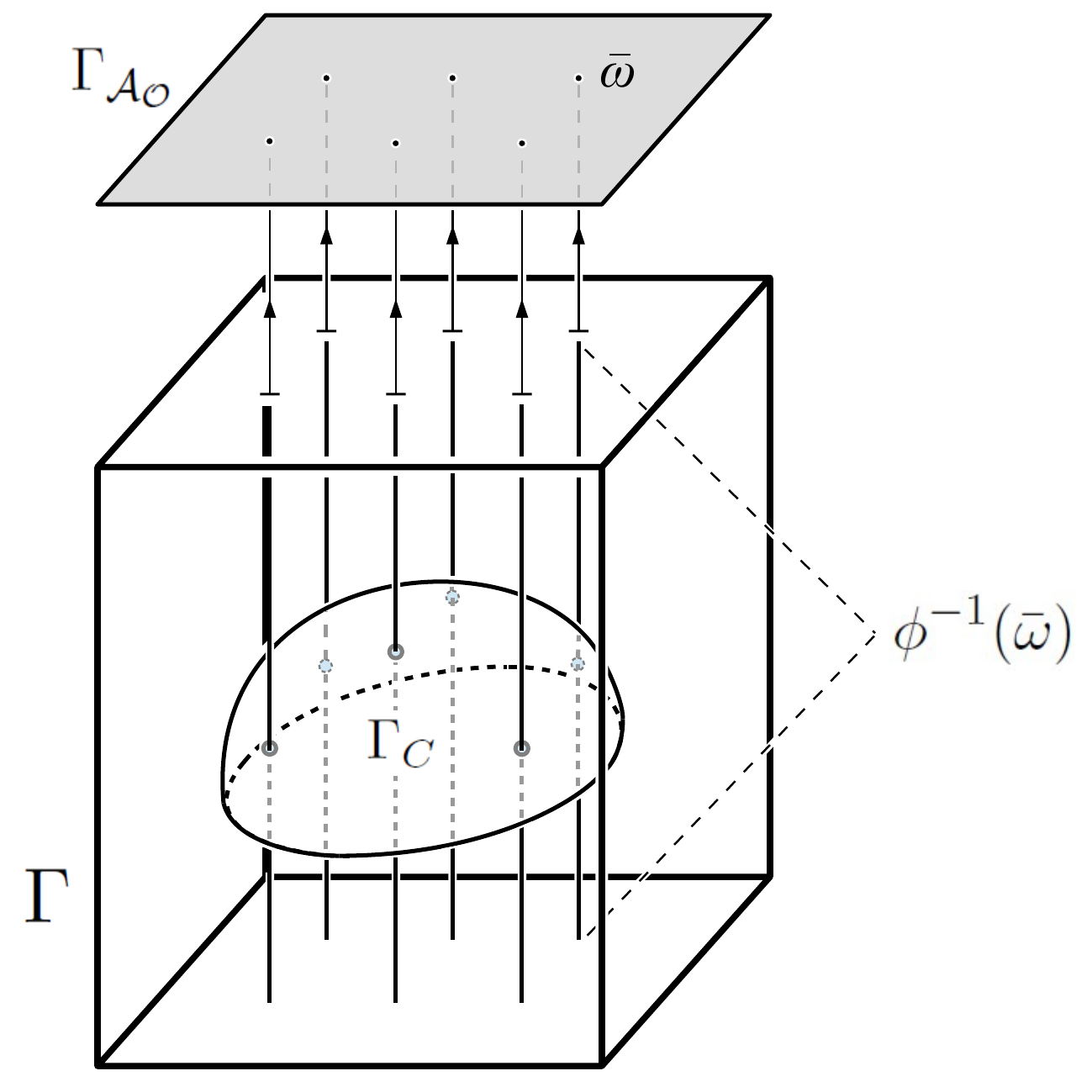}
  \caption{Schematic representation of the fibers of points on
      $\Gamma_{\mathcal{A}_{\mathcal{O}}}$\ under the map $\phi$. The box
      represents all of $\Gamma$\ with $\Gamma_C$\ represented by the curved
      surface inside. The shaded plane outside of the box represents
      $\Gamma_{\mathcal{A}_{\mathcal{O}}}$\ with vertical lines inside the box
      representing fibers in $\Gamma$: the elements of each fiber are all
      mapped to a single point in $\Gamma_{\mathcal{A}_{\mathcal{O}}}$. If
      each fiber intersects $\Gamma_C$\ at a single point, then there is a
      one-to-one correspondence between states in
      $\Gamma_{\mathcal{A}_{\mathcal{O}}}$\ and those in
      $\Gamma_C$. }\label{fig:Fibers}
\end{center}
  \end{figure}

To set up our new procedure, let
$\mathcal{A}_{\mathcal{O}} \subset \mathcal{A}$\ be some unital
$*$--subalgebra, and let $\Gamma_{\mathcal{A}_{\mathcal{O}}}$\ denote
normalized linear states on $\mathcal{A}_{\mathcal{O}}$, defined by
restricting the original $\Gamma$ to the subalgebra. In this way, any state on
$\Gamma$, including those on the constraint surface $\Gamma_C$, can be projected
by $\phi \colon \Gamma \rightarrow \Gamma_{\mathcal{A}_{\mathcal{O}}}$, where
$\phi(\omega)(B) :=\omega(B)$\ for
$B\in \mathcal{A}_{\mathcal{O}}$.\footnote{This projection is the pullback of
  linear functionals under the inclusion
  $\imath\colon \mathcal{A}_{\mathcal{O}} \hookrightarrow \mathcal{A}$, so that
  $\phi(\omega) = \omega \circ \imath$.}  In order to make constraint
solutions (and, later, also gauge-fixed states) interpretable as states on a
$*$--algebra, we would like to use $\Gamma_{\mathcal{A}_{\mathcal{O}}}$\ to
represent $\Gamma_C$ in some way. To see when this is possible, consider the fiber of a
state $\bar{\omega} \in \Gamma_{\mathcal{A}_{\mathcal{O}}}$\ under the map
$\phi$, schematically represented in Figure~\ref{fig:Fibers},
\[
\phi^{-1}(\bar{\omega}) := \{ \omega \in \Gamma : \omega(B) = \bar{\omega}(B)
\;\mbox{for all}\; B\in \mathcal{A}_{\mathcal{O}} \} \ .
\]
Given a state $\bar{\omega} \in \Gamma_{\mathcal{A}_{\mathcal{O}}}$, the
corresponding fiber has a non-zero intersection with $\Gamma_C$\ if there is a
state $\omega \in \phi^{-1}(\bar{\omega})$\ such that $\omega(AC) = 0$\ for
all $A \in \mathcal{A}$. This implies that for all
$B \in \mathcal{A}_{\mathcal{O}} \cap \mathcal{A}C$\ we have
$\bar{\omega} (B):=\omega(B) = 0$. Therefore, in order to be able to identify
\emph{every} state in $\Gamma_{\mathcal{A}_{\mathcal{O}}}$\ with a non-empty a
region of $\Gamma_C$, we need to have
$\mathcal{A}_{\mathcal{O}} \cap \mathcal{A}C=\{0\}$. Under what circumstances
does a state in $\Gamma_{\mathcal{A}_{\mathcal{O}}}$\ represent only a
\emph{single} state on $\Gamma_C$? This happens if the value a state assigns
to the subalgebras $\mathcal{A}_{\mathcal{O}}$\ and $\mathcal{A}C$\ linearly
extends to the whole of $\mathcal{A}$, i.e. if linear combinations of elements
in $\mathcal{A}_{\mathcal{O}} \cup \mathcal{A}C$\ span the whole of
$\mathcal{A}$.

Let us suppose that we are in possession of a $*$-subalgebra
$\mathcal{A_O}\subset\mathcal{A}$\ such that
$\mathcal{A}=\mathcal{A_O} + \mathcal{A}C$\ and
$\mathcal{A}_{\mathcal{O}} \cap \mathcal{A}C=\{0\}$, so that there is a
one-to-one map between solutions to the constraint and states on
$\mathcal{A_O}$ and $\phi$ is invertible. We can now analyze the gauge orbits
by mapping them from $\Gamma_C$\ to
$\Gamma_{\mathcal{A}_{\mathcal{O}}}$. Indeed, any curve $\omega_{\lambda}$\ on
$\Gamma_C$ can be mapped to a curve $\bar{\omega}_{\lambda}$\ on
$\Gamma_{\mathcal{A}_{\mathcal{O}}}$, via
$\bar{\omega}_{\lambda}(B) := \omega_{\lambda}(B)$\ for all
$B \in \mathcal{A}_{\mathcal{O}}$.

In order to characterize the orbits corresponding to physical states we will
fix gauge freedom by imposing a sufficient number of additional algebraic
conditions on the states, $\bar{\omega}(A)=0$\ for all
$A\in \mathcal{J}\subset\mathcal{A_O}$ with a suitable $\mathcal{J}$.  If
$\mathcal{J}$\ is a linear subspace, then we can construct the linear quotient
space, denoting the canonical map
$\pi\colon\mathcal{A_O} \rightarrow \mathcal{A_O}/\mathcal{J}$. furthermore, if
$\mathcal{J}$\ is a $*$-ideal of $\mathcal{A_O}$, then
$\mathcal{A_O}/\mathcal{J}$\ inherits the structure of a $*$-algebra from
$\mathcal{A_O}$ (and hence from the kinematical algebra
$\mathcal{A}$). Geometrically, the normalized states on
$\mathcal{A_O}/\mathcal{J}$\ can be pulled back to $\Gamma_{\mathcal{A_O}}$\
corresponding precisely to the states $\bar{\omega}$\ that annihilate
$\mathcal{J} = \ker \pi$, we denote this gauge-fixing surface
$\left. \Gamma_{\mathcal{A_O}} \right|_{\pi} :=\{ \bar{\omega} \in
\Gamma_{\mathcal{A_O}} : \bar{\omega}(B) = 0 \;\mbox{for all}\; B \in\ker \pi \}$.

Each state on the quotient algebra $ \mathcal{A_O}/\mathcal{J}$\ can be pulled
back to a unique state on $\Gamma_{\mathcal{A_O}}$ (belonging to the
gauge-fixing surface), which can be further mapped to a unique state on the
constraint surface $\Gamma_C$\ by inverting the map $\phi$. We have therefore
selected a subset of solutions to the quantum constraint that can be
interpreted as states on a different unconstrained algebra, representing the
degrees of freedom that remain after gauge-fixing. Revisiting our desiderata
from the start of this section, we still need to determine whether the
entirety of gauge freedom has been fixed (desideratum (i)), whether all of the
physical states are sampled (desideratum (ii)), and whether an analog of
positivity on $\mathcal{A}_{\rm obs}$\ can be imposed in some
way. Investigating these properties further requires additional assumptions
about the kinematical algebra and the constraint element.

\subsection{Relational gauge}\label{sec:RelationalGauge}

The construction schematically described in the previous subsection can be
concretely realized provided the constrained system possesses a suitable
reference observable $Z=Z^* \in \mathcal{A}$. Gauge fixing will be
accomplished by fixing the value of the reference observable; the remaining
freedom is characterized by observables that can be specified simultaneously
with $Z$\ and are therefore part of its commutant
$Z'=\{A\in \mathcal{A} : [A,Z] = 0\}$. By analogy with $\mathcal{A}_{\rm obs}$, $Z'$\ is a unital $*$--subalgebra of $\mathcal{A}$, which will play the role of $\mathcal{A}_{\mathcal{O}}$\ from the previous section. As discussed there, in
order for states on $Z'$\ to be in one-to-one correspondence with the
constraint surface $\Gamma_C$, we additionally require
$Z'+\mathcal{A}C = \mathcal{A}$\ and $Z'\cap\mathcal{A}C = \{0\}$. What freedom within $Z'$\ could we use to define gauge--fixing conditions? The commutant of $Z$ always
contains $Z$ itself, which, as in a deparameterization procedure, would no
longer be considered a (non-trivial) physical observable once it has been
identified with a time parameter during the gauge--fixing process. The algebra of
observables that are accessible once a choice of $Z$ as reference observable
has been made must therefore derive from $Z'$\ in some way that
eliminates $Z$.

An algebra element can be eliminated from its commutant by
factoring out the ideal defined by the element times the commutant, here
$ZZ'$, such that $[Z]=[0]$ for equivalence classes in the quotient space. More
generally, any central element can be added to $Z$ in this product. We make
use of this freedom in order to simultaneously eliminate $Z$ from the remaining observables and fix its value, by considering the one-parameter family of quotient algebras
$Z'/(Z-t\mathbf{1})Z'$, for $t\in \mathbb{R}$. In these quotient spaces, $[Z-t\mathbf{1}]=[0]$ or
$[Z]=t[\mathbf{1}]$. (Even though the equivalence class $[A]$\ depends on the choice of $t$, we omit reference to $t$\ in order to reduce notational clutter.) More generally $[A] = [B]$\ if $(A-B)$\ is in the $*$--ideal $(Z-t\mathbf{1})Z'$, which will play the role of $\mathcal{J}$ introduced at the end of section~\ref{sec:Geometry}. Denoting the natural projection $\pi_t\colon Z' \rightarrow Z'/(Z-t\mathbf{1})Z'$, where $\pi_t(A)=[A]$, we see that the kernel of this map is precisely the ideal $ \ker \pi_t= (Z-t\mathbf{1})Z'$, since $[(Z-t\mathbf{1}) A] = [0]$\ for any $A$\ in  $Z'$. 

The quotient space $Z'/(Z-t\mathbf{1})Z'$ is always
a $*$-algebra under structure inherited from $\mathcal{A}$, and it is unital unless $Z\propto\mathbf{1}$. However it is not a subalgebra of $Z'$ (or of $\mathcal{A}$) and therefore cannot
directly define the algebra of ``system observables other than $Z$.''  Instead,
copying the language of~\cite{AlgebraicTime}, we define that a subalgebra
$\mathcal{F}\subset Z'$ is a {\bf fashionable algebra} compatible with $Z$ if
it is a unital $*$-algebra such that for all $t \in \mathbb{R}$ we have
$\mathcal{F} \cap \ker \pi_t = \{ 0\}$. Therefore,
$\pi_t(\mathcal{F})=Z'/(Z-t\mathbf{1})Z'$ is an isomorphism for each value of $t$.  A {\bf quantum clock} $(Z, \mathcal{F})$\ is a reference
observable $Z=Z^* \in \mathcal{A}$\ together with a compatible fashionable
algebra ${\cal F} \subset Z'$.\footnote{The notation of \cite{AlgebraicTime} was
developed for a clock that in general may be valid only in a finite range and
may have to be transformed once the limits are reached. The
label $Z$ therefore signifies a ``Zeitgeist'' and fashionables in
$\mathcal{F}$ fall out of fashion once the Zeitgeist changes because a new
clock must be introduced.} Here, as in~\cite{AlgebraicTime}, our ultimate goal is to use $Z$\ to track the
passage of time, hence the term ``clock'', which we will also use to refer to
$Z$\ itself. However, this construction can also be used to fix non-temporal
gauge freedoms (e.g. associated with translation invariance). As we will
discuss in section~\ref{sec:LinearCGauge}, the fashionable algebra $\mathcal{F}$\ is used to
relate the states corresponding to different values of $Z$, such that we can
uniquely identify the evolution of a given observable with respect to $Z$.
Such a relationship is critical when fixing a temporal gauge, but may not be
important for constraints generating non-temporal transformations. For the
rest of this subsection we will focus on the properties of a single gauge slice
(fixed $t$) and mostly ignore $\mathcal{F}$.

 Given that our clock satisfies all of the above requirements, the constraint surface $\Gamma_C$\ maps to $\Gamma_{Z'}$, and  fixing $Z$\ to a value $t$\ defines a
gauge-fixing surface
\begin{equation} \label{gaugefixing}
  \left. \Gamma_{Z'} \right|_{\pi_t}=\{ \bar{\omega} \in
\Gamma_{Z'} : \bar{\omega}(B) = 0 \;\mbox{for all}\; B \in\ker \pi_t \} \subset \Gamma_{Z'} \ .
\end{equation}
Under what circumstances does this surface fix all of the gauge freedom?
According to the analysis of Section~3.5 of~\cite{AlgebraicTime}, the additional
condition required in order for this gauge choice to fix all of the flows
generated by the elements of $\mathcal{A}C$, at least locally, is that for
this value of $t$,
\[
[Z, C] = i a_t \mathbf{1} + (Z-t\mathbf{1})B_t \ ,
\]
for some $a_t\in \mathbb{R}$, and some $B_t \in \mathcal{A}$, such that
$[B_t, A] = 0$\ for all $A\in Z'$. A simple way to ensure that this relation
holds for a continuous range of values of $t$, as necessary for tracking time
evolution relative to $Z$, while also fixing the value of $a_t$,  is to demand a stronger condition,
\[
[Z, C] = i\hbar \mathbf{1} \ ,
\]
which we will use in the
sequel.\footnote{In fact the proof of local gauge fixing Section~3.2
  of~\cite{AlgebraicTime} explicitly assumes the stronger condition
  $[Z, C] = i\hbar \mathbf{1}$, however all steps of the proof carry over with
  minor changes if the weaker condition
  $[Z, C] = i\hbar a_t \mathbf{1} + (Z-t\mathbf{1})B_t$, with
  $[B_t, Z'] = \{0\}$\ is used instead.} This condition ensures that the gauge
is completely fixed in the sense that no flows generated by the adjoint action
of $\mathcal{A}C$\ are tangential to the gauge-fixing surface
$\left. \Gamma_{Z'} \right|_{\pi_t}$.

Revisiting our desiderata from Section~\ref{sec:Geometry}, gauge-fixing
associated with specifying the value of a reference observable $Z$\ that is
canonically conjugate to the constraint, satisfies a local version of
desideratum (i), guaranteeing that each point on the gauge-fixing surface
$\left. \Gamma_{Z'} \right|_{\pi_t}$\ has some open neighborhood where the
same physical state is not represented by any other state on the surface. It
is still possible that a given physical state corresponds to two or more
distinct gauge-fixed states globally. Desideratum (ii) is explicitly of a
global nature and we do not have a result that guarantees that every physical
state corresponds to some state on $Z'/(Z-t\mathbf{1})Z'$ (see discussion in
Section~3.5 of~\cite{AlgebraicTime}). Desideratum (iii) is fully satisfied,
since each gauge-fixed state is a state on $Z'/(Z-t\mathbf{1})Z'$. Moreover a
state that is positive on $Z'/(Z-t\mathbf{1})Z'$\ with respect to the
inherited $*$-structure corresponds to a solution of the constraint that is
positive on $\mathcal{A}_{\rm obs}$ (Lemma 13 of~\cite{AlgebraicTime}).

\subsection{Dynamical reduction of a Hamiltonian constraint} \label{sec:LinearCGauge}

We will now focus on the case of a completely constrained system, where the
constraint also plays the role of the Hamiltonian. In addition to correctly
removing constrained degrees of freedom, the major objective of implementing a
Hamiltonian constraint is to address the problem of dynamically interpreting
the constrained system, as is directly relevant to canonical attempts of
quantizing gravity. We refer to any procedure that implements this additional
requirement, compared with constrained systems which have an unconstrained
Hamiltonian, as {\bf \emph{dynamical} symplectic reduction.}

Our efforts toward a quantization of dynamical symplectic reduction are guided
by the intuition we gain from the process of parameterization of an ordinary
quantum mechanical system. Starting with a Hamiltonian system with evolution
of states given by
\begin{equation}\label{eq:Hflow}
  \frac{{\rm d}\omega_t(A)}{{\rm d} t} = \frac{1}{i\hbar} \omega_t \left(
    [A,H] \right) + \omega_t \left( \frac{\rm d}{{\rm d} t} A \right) \ ,
\end{equation}
generated by $H\in {\cal A}$, one can formally extend the kinematical algebra
by two new generators $T=T^*$ (time) and $E=E^*$ (energy) with
$[T,E]=i\hbar \mathbf{1}$ and $[T,A]=0=[E,A]$ for all $A\in\mathcal{A}$. On
this extended algebra $\mathcal{A}_{\rm ext}$\ the Hamiltonian is replaced by
a constraint $C:=E+H$. Schematically, recovering the original dynamical system
from the constrained system with the extended algebra of degrees of freedom,
$\mathcal{A}_{\rm ext}$, is accomplished in two steps. First, the generator $E$\
is eliminated by solving the constraint itself: Since $E=C-H$, the action of
$E$\ on solutions to the constraint is equivalent to that of $-H$. Second, we
look at the flow generated by the constraint on the remaining degrees of
freedom ($\mathcal{A}_{\rm ext}$\ sans elements constructed using $E$)
\begin{eqnarray} \label{eq:OCflow} 
\frac{1}{i\hbar} \omega_t \left([A,C] \right) &=& \frac{1}{i\hbar}
                                                  \omega_t\left( [A,H] + [A,E] \right)\nonumber\\
  &=& \frac{1}{i\hbar}\omega_t \left( [A,H]
\right) + \omega_t \left( \frac{\rm d}{{\rm  d} T} A \right)  \ , 
\end{eqnarray}
where we allow $A$ to explicitly depend on $T$\ polynomially. (In this case,
the derivative by an algebra element $T$ is defined by reordering terms in the
commutator $[A,E]$ and does not require the introduction of an operator
topology on the algebra.)

Although the resulting equation looks very similar to~(\ref{eq:Hflow}), the
flow~(\ref{eq:OCflow}) is equivalent to the original Hamiltonian flow only
when $T$ is formally demoted to a parameter, rather than an element of the
kinematical algebra, and identified with $t$. Intuitively, this reverse of a
parameterization process of passing from the constrained system with algebra
$\mathcal{A}_{\rm ext}$\ back to the dynamical system with a smaller algebra
$\mathcal{A}$\ (or \emph{deparameterization}) can be interpreted as fixing the
observable $T$\ to take a specific value of the parameter $t$. The values
assigned by dynamical states of the original unconstrained system at time $t$,
solving (\ref{eq:Hflow}), can be constructed using the corresponding physical
states of the constraint $C$, by restricting the values they assign to $T$\
and other elements of $\mathcal{A}_{\rm ext}$ derived from it, and inserting
these values in the flow (\ref{eq:OCflow}). The main result
of~\cite{AlgebraicTime} is the definition of a generalized algebraic version of
this process, which we summarize in the rest of this subsection.

Gathering the conditions identified in Sections~\ref{sec:Geometry} and
~\ref{sec:RelationalGauge}, we define the algebraic version of
deparameterization: A quantum constraint $C\in \mathcal{A}$\ is {\bf
  deparameterized} by the clock $(Z, \mathcal{F})$\ if
$[Z, C] = i\hbar \mathbf{1}$\ and the commutant of $Z$\ is such that (1)
$Z'\cap \mathcal{A}C = \{0\}$, and (2) the set $Z'\cup \{C\}$\ algebraically
generates $\mathcal{A}$.  It is straightforward to verify that conditions 1
and 2 together imply, in addition, $Z'+\mathcal{A}C = \mathcal{A}$\ as
required for a one-to-one mapping between the states on $Z'$\ and solutions to
the constraint. As already discussed in Section~\ref{sec:Geometry}, for each
value $t\in\mathbb{R}$, the positive states on the quotient $*$-algebra
$Z'/(Z-t\mathbf{1})Z'$\ correspond to physical states on the observable
algebra. The flow associated with the adjoint action of the Hamiltonian
constraint itself is one of the gauge freedoms fixed by choosing a value of
$Z$.

Since, by our definition of a deparameterized quantum constraint, the clock
and the constraint are canonically conjugate, one can intuitively expect this
flow to ``evolve the clock'' and take a gauge-fixed state from one value of
the reference observable $Z$ to another. To see that this is indeed the case,
we first note that the adjoint action of $C$ preserves the subalgebra $Z'$,
since for any $A\in Z'$
\[
\left[ \left[A, C\right], Z\right] = \left[ \left[Z, C\right], A\right]
+\left[ \left[A, Z\right], C\right] = 0 \ , 
\]
using $[Z, C]=i\hbar \mathbf{1}$ and $[A, Z]=0$. Therefore $[A,C]\in Z'$, so
that $[\cdot,C]$\ defines a flow $S_{C}(\lambda)$ on $\Gamma_{Z'}$\ via
\begin{equation} \label{eq:TimeFlow}
i\hbar \frac{{\rm d}}{{\rm d} \lambda} \left( S_{C}(\lambda) \omega(A) \right) :=  S_{C}
(\lambda) \omega([A, C])  \ \  \mbox{and}\ \ S_{C}(0) = {\rm id}\ . 
\end{equation}
A state in which the value of $Z$\ is fixed to some $t$\ is a state on the
quotient algebra $Z'/(Z-t\mathbf{1})Z'$\ and can be pulled back to some state
on the gauge-fixing surface
$\bar{\omega} \in \left. \Gamma_{Z'} \right|_{\pi_t}$. Lemma 10
of~\cite{AlgebraicTime} shows that in this case
$S_{C}(\lambda) \bar{\omega} \in \left. \Gamma_{Z'}
\right|_{\pi_{t+\lambda}}$, which corresponds to a state on the quotient
algebra $Z'/(Z-(t+\lambda)\mathbf{1})Z'$. Furthermore, since $C^*=C$, this
flow preserves positivity of states on $\mathcal{A}$\ (see Lemma 2
of~\cite{AlgebraicTime}) and, by restriction, also of states on $Z'$. Since
$S_{C}(\lambda)$\ is a gauge flow, $\bar{\omega}$\ and
$S_{C}(\lambda) \bar{\omega}$\ correspond to the same physical state. The flow
$S_{C}$\ can now be interpreted as time evolution relative to the clock $Z$:
\begin{itemize}
\item Given an observable $A=A^*$ such that $[A, Z] = 0$, we can say that
  ``$A$\ when $Z=t$'' corresponds to the element of the quotient algebra
  $\pi_t(A) \in Z'/(Z-t\mathbf{1})Z'$.
\item Knowing the relational state
  $\bar{\omega}_{t_1}\in\left. \Gamma_{Z'} \right|_{\pi_{t_1}}$ of the
  constrained system at some time $t_1$\ corresponds to knowing all the values
  assigned to such relational observables ``at $Z=t_1$'';
\item We can deduce the values assigned by the same physical state at a
  different value of the clock $t_2$\ by integrating the flow of
  equation~(\ref{eq:TimeFlow}) for a duration of $t_2-t_1$.
\end{itemize}
 
One more hurdle remains before we can cast the constrained system as some
unconstrained system evolving ``in time $Z$'': We have to specify exactly which
degrees of freedom are evolving relative to $Z$. The flow described by
equation~(\ref{eq:TimeFlow}) acts on the states on the commutant $Z'$\ that
comprise $\Gamma_{Z'}$, while the relational observables ``$A$\ when $Z=t$''
are elements of the quotients $Z'/(Z-t\mathbf{1})Z'$, which are different
algebras for different values of $t$\ with no canonical mapping between
them. We could propose that the reduced system is described by the entire set
of observables that can be specified simultaneously with $Z$, which comprise
$Z'$, evolving in relation to $Z$. However in regards to our gauge fixing
procedure, $Z'$ is over-complete: its elements cannot all be assigned values
independently by a gauge fixed state at some $t$---two observables that differ
by $(Z-t\mathbf{1})G$, for some $G\in Z'$\ will be assigned identical values
by any state in which the value of $Z$\ is fixed to $t$.

This place is precisely where the fashionable algebra introduced in our
definition of a quantum clock becomes important. From this definition it
follows that, for each $t$, the canonical projection $\pi_t$\ restricted to
$\mathcal{F}$\ is a $*$-isomorphism (see discussion in Section~2.4
of~\cite{AlgebraicTime}). The fashionable algebra can therefore be used to
characterize the degrees of freedom at every constant value of the clock
$Z$. Using a geometrical analogy, elements of $\mathcal{F}$\ serve as local
coordinates along the gauge-fixed surfaces, while the clock $Z$\ provides
coordinates in the normal directions. An initial state can be freely specified
on $\mathcal{F}$, and will evolve in time $Z$\ along the flow
$S_{C}(\lambda)$. In other words, the constrained system, when
deparameterized, is equivalent to an unconstrained system with degrees of
freedom given by $\mathcal{F}$\ and a time evolution flow. Schematically, for
every value $t$\ taken by the clock, deparameterization requires the
kinematical algebra to decompose into subalgebras that share only the null
element,
\[
\mathcal{A} =\mathcal{A}C + (Z-t\mathbf{1})Z' + \mathcal{F} \ ,
\]
where $\mathcal{F}$ is a $*$--subalgebra of $Z'$ isomorphic to
$Z'/(Z-t\mathbf{1})Z'$ at each $t$. We make use of this decomposition to
streamline the usage of different spaces of states:
A state $\omega \in \Gamma$ is {\bf almost-positive} with respect to a
deparameterization of $C$ by $Z$ if 
\begin{enumerate}
\item it annihilates the left ideal generated by $C$: $\omega(AC)
  = 0$ for all $A \in \mathcal{A}$; 
\item it is positive on the commutant of $Z$: $\omega(BB^*) \geq 0$
  for all $B \in Z'$; 
\item it parameterizes left multiplication by $Z$: for all
  $A \in \mathcal{A}$, $\omega(ZA) = \omega(Z) \omega(A)$.  
\end{enumerate}

Because the three component algebras of $\mathcal{A}$\ share only the null
element, the three conditions of the definition of an almost-positive state
can be imposed independently. The first condition ensures that an
almost-positive $\omega$ is a solution of the constraint. The second condition
ensures that $\omega$ restricts to some positive state on $Z'$, and hence also
on $\mathcal{F}$. The third condition ensures that this restriction belongs to
the constant clock surface $\left. \Gamma_{Z'}\right|_{\pi_{\omega(Z)}}$.

In other words, an almost-positive state $\omega$, corresponds to a
gauge-fixed state where the value of $Z$\ is fixed to $\omega(Z)$, which is
also a positive state on $\mathcal{F}$\ by restriction. The converse is also
true: given a value $t\in \mathbb{R}$\ of the clock and positive state $\tilde{\omega}$\ on
$\mathcal{F}$, there is a unique almost-positive state $\omega$, such that
$\omega(Z)=t$\ and $\left. \omega \right|_{\mathcal{F}}=\tilde{\omega}$ (see
Lemma 7 and Corollary 2 in~\cite{AlgebraicTime}). Furthermore, as expected, the
flow $S_{C}(\lambda)$\ preserves almost-positivity (Lemma 12
in~\cite{AlgebraicTime}). Once we have identified the appropriate algebras, the
particular solutions to deparameterized dynamics can proceed via
almost-positive states on the original kinematical algebra. These states may
also be useful if one is looking for a way to relate different clock choices:
unlike the states on quotients of different commutants, almost-positive states
associated with different clocks all exist on the same space of states---the
original space of kinematical states $\Gamma$. (A full treatment of clock
changes in the algebraic formulation remains to be completed.) 

\subsection{Algebraic deparameterization with a canonical clock}
\label{sec:LinearConstraintAlgebra}

The conditions for deparameterization essentially require the clock degree of
freedom to form a canonical pair with the Hamiltonian constraint
$[Z, C] = i\hbar \mathbf{1}$. If our algebra has a canonical decomposition,
generated by $Z$, $C$ and some $\{ Q_1, Q_2, \ldots; P_1, P_2, \ldots \}$\
with $[ Q_i, P_j ] = i\hbar \delta_{ij} \mathbf{1}$\ and
$[C, Q_i] = [C, P_i] = [Z, Q_i] = [Z,P_i] = 0$, then we already have the
solution to the constraint in the form of the physical observables $Q_i$\ and
$P_i$. This is not the usual situation. In this subsection, we apply algebraic
deparameterization to the more common scenario where the clock variable has a
known kinematical conjugate $E=E^*$ not equal to $C$, such that
$[Z, E] = i \hbar \mathbf{1}$ while $Z$\ and $E$\ commute with all elements of
$\mathcal{A}$\ that are independent of both $Z$\ and $E$.

\subsubsection{Canonical tensor product algebra}

If the constraint in such a system is deparameterizable by $Z$ according to
our definition, it follows that $E=AC + B$\ for some $A\in \mathcal{A}$\ and
some $B \in Z'$. The canonical commutation relation between $Z$\ and $E$\ then
implies $A = \mathbf{1}$, and, setting $H=-B$, we can write the constraint as
\begin{equation} \label{eq:ParamParticleC}
C=E+H \ .
\end{equation}
The condition $C=C^*$\ immediately implies $H=H^*$, and $[H, Z]=0$\ follows
since $H=-B\in Z'$. We will refer to this type of constrained system as
\emph{parameterized particle}, since it has precisely the form that leads to
equation~(\ref{eq:Hflow}).

In this scenario, the kinematical algebra has the structure of a tensor
product of two algebras,
$\mathcal{A} \cong \mathcal{A}_T \otimes \mathcal{A}_S$. We will denote the
corresponding *-algebra isomorphism
$\Phi\colon \mathcal{A} \rightarrow\mathcal{A}_T \otimes \mathcal{A}_S$.  The
first component consists of complex polynomials over a canonical pair,
$\mathcal{A}_T = \mathbb{C}[\tau, \epsilon]$\ with
$(\tau \epsilon - \epsilon \tau) = i\hbar \mathbf{1}_T$---it keeps the
time. We have $\Phi(Z) = \tau \otimes \mathbf{1}_S$\ and
$\Phi(E) = \epsilon \otimes \mathbf{1}_S$. The
second component corresponds to the degrees of freedom of the``rest of the
system" that evolve in $Z$---ones that can be specified simultaneously with
either $E$\ or $Z$. The ``system'' component is precisely the commutant of
the ``clock'' component (but not necessarily vice versa, since the system
component may contain a non-trivial center)
\[
\Phi^{-1}\left(\mathbf{1} \otimes \mathcal{A}_S \right) = \{ A \in \mathcal{A}
: [Z, A] = [E, A] = 0 \} \ . 
\]

The two components share the null and identity elements. The commutant of the
clock variable is the subalgebra generated by polynomials in $Z$\ and elements
of the system algebra:
$Z'=\Phi^{-1} \left( \mathbb{C}[\tau] \otimes \mathcal{A}_S \right)$. When the
clock is part of a canonical subsystem, as described here, there is an obvious
choice for the fashionable algebra
$\mathcal{F} = \Phi^{-1} \left( \mathbf{1} \otimes \mathcal{A}_S \right)$,
which has the property that $[\mathcal{F}, E]=\{ 0 \}$. This latter property
is not necessary for deparameterization as defined here, but it allows us to
interpret $E$\ as the generator of \emph{time translation}: The flow that it
generates through the commutator increases the value a state assigns to $Z$,
while keeping the values of the fashionables fixed.

Let us further assume that the ``system" algebra
$\mathcal{F} = \Phi^{-1} \left( \mathbf{1} \otimes \mathcal{A}_S \right)$\ has
a linear basis $\{A_i\}_{i\in I}$\ where $I$\ is some set of indices, and
spell out in a bit more detail what deparameterization looks like in this
case. The full kinematical algebra $\mathcal{A}$\ is then spanned by elements
$\{A_iZ^mE^n\}$. (Differently ordered products of powers of $Z$\ and $E$ of
order $N=m+n$\ can be re-ordered using the canonical commutation relation by
adding terms of order $(N-1)$\ or less.) Basis elements $A_i$\ will, in
general, have non-vanishing commutators among themselves, and we have
\[
[Z, E] = i\hbar \mathbf{1} \ , \quad [Z, A_i]=[E, A_i] = 0 \ .
\]
The commutant $Z'$ of the clock is then spanned by $\{A_iZ^m\}$, which is a
subset of the basis for $\mathcal{A}$, with $n=0$. For any $t\in \mathbb{R}$,
the ideal $(Z-t\mathbf{1})Z'$\ defines cosets $[A] \subset Z'$\ for each
$A\in Z'$, where $\tilde{A} \in [A]$ if $(A-\tilde{A}) = (Z-t\mathbf{1})B$\
for some $B \in Z'$. In particular, for a basis element
\begin{eqnarray*}
A_i Z^m &=&  A_i  \left( \left( Z -t\mathbf{1} \right) +t\mathbf{1}\right)^m \\
 &=& A_i \sum_{k=0}^{m} {m\choose k}  \left( Z -t\mathbf{1} \right)^k t^{m-k}\\
 &=&  A_i t^m + A_i   \sum_{k=1}^{m} {m\choose k}  \left( Z -t\mathbf{1} \right)^k t^{m-k} \ .
\end{eqnarray*}
The sum at the end of the final expression lies in the ideal
$(Z-t\mathbf{1})Z'$, and, therefore in the coset of the zero element; hence,
$ \left[ A_i Z^m \right] = \left[ A_i t^m \right]$.

The collection of cosets
$\left\{ \left[ A_i \right] \right\}_{i\in I}$, therefore, provides a linear
basis on the quotient algebra $Z'/(Z-t\mathbf{1})Z'$. In terms of this basis,
the action of the canonical projection
$\pi_t\colon Z' \rightarrow Z'/(Z-t\mathbf{1})Z'$ on a basis elements of $Z'$\ is
\[
\pi_t \left( A_i  Z^m \right) = \pi_t \left( A_i  t^m \right) = t^m \left[  A_i  \right] \ ,
\]
which extends to arbitrary $A\in Z'$\ by linearity. Noting that
$\{A_i\}_{i\in I}$\ is precisely the linear basis on $\mathcal{F}$\ that we
started out with, we see that $\pi_t$\ restricted to $\mathcal{F}$ (setting
$m=0$) is an isomorphism. This immediately implies that $\mathcal{F}$\
satisfies the two conditions, $\mathcal{F} \cap \ker \pi_t = \{ 0\}$ and
$\pi_t(\mathcal{F}) = Z'/(Z-t\mathbf{1})Z'$, required for a fashionable
algebra. For each $t$, any element of $Z'$\ can be canonically projected to
$Z'/(Z-t\mathbf{1})Z'$\ and then mapped to $\mathcal{F}$, which gives us a
one-parameter family of projections $\alpha_t\colon Z' \rightarrow \mathcal{F}$
\begin{equation}\label{eq:FProjection}
\alpha_t \left( A_i  Z^m \right) = t^m  A_i  \ .
\end{equation}
Each fashionable is projected to itself, while the projections of other
elements of $Z'$\ to fashionables are $t$-dependent.

\subsubsection{Almost-positive states}
\label{sec:AlmostPositive}

Let us now examine how the conditions of deparameterization come into play in
the above scenario. We have already assumed that $[Z, C] = i\hbar \mathbf{1}$; we will now verify that other requirements of deparameterization laid out in section~\ref{sec:LinearCGauge} are also satisfied. We
note that any non-zero element of $\mathcal{A}C$\ contains at least one term
with a factor of $E$. When written in terms of the basis elements, it then
contains at least one term $A_iZ^mE^n$ with $n\neq0$. Therefore,
$Z'\cap \mathcal{A}C=0$, as required for deparameterization. We now write $E=C - H$, so
that our basis elements can be re-written
\begin{eqnarray*}
A_iZ^mE &=& - A_iZ^m H+ A_iZ^m C\\
A_iZ^mE^2 &=& \left( H^2 +[H, C] \right) -2HC+C^2 
\end{eqnarray*}
Iterating this process and recalling from Section~\ref{sec:LinearCGauge} that
$[\cdot,C]$\ maps $Z'$\ to $Z'$, we have
\begin{eqnarray}\label{eq:CBasis}
A_iZ^mE^n &=& A_iZ^m \sum_{j=0}^n B_{nj}\, C^j \ ,
\end{eqnarray}
for some $B_{nj}\in Z'$. Clearly, our chosen basis of $\mathcal{A}$\ and,
therefore, also $\mathcal{A}$\ itself is algebraically generated by
$Z'\cup \{C\}$\ as required.

Now, dynamical evolution in $Z$\ is constructed by passing to almost-positive
states with respect to this deparameterization of $C$ and focusing on the flow
that is generated on those states by $C$. Evaluating a basis element in an
almost-positive state using expression~(\ref{eq:CBasis}) and condition 1 of
the definition of an almost-positive state, we get a collection of relations, one for each basis element:
\[
\omega(A_iZ^mE^n) = \omega(A_iZ^m B_{n0}) \ .
\]
One way to interpret these relations is that they place no restrictions on the
values an almost-positive state assigns to $Z'$ (spanned by $\{ A_iZ^m \}$),
but once those values are set, the above relations uniquely extend the state
to the rest of $\mathcal{A}$ (spanned by $\{A_iZ^mE^n \}$). Condition 3 of the
definition is linear and can therefore also be sufficiently satisfied if
imposed on the values of basis elements: this time it will be the basis of
$Z'$, rather than the full kinematical algebra. We get
\[
\omega(A_iZ^m) = \omega(Z)^m \omega(A_i) \ .
\]

Thus, the values assigned by almost-positive states to all elements of
$\mathcal{A}$\ can be derived from the expectation value of the clock
$\omega(Z)$\ and the values they assign to the fashionables. In fact,
expectation values on $Z'$\ in almost-positive states can be taken after
projecting to fashionables using~(\ref{eq:FProjection}). It is not difficult
to verify that for $A\in Z'$\ and any almost-positive state
$\omega ( A) = \omega \left( \alpha_{\omega(Z)} (A) \right)$ with $\alpha_t$
defined in (\ref{eq:FProjection}). We will shortly
use this property to project time evolution to fashionables.

Applying condition 2 of the definition of almost-positive states to basis
elements of $Z'$ we get
\[
\omega(A_iZ^{2m} A_i^*) = \left( \omega(Z)^2\right)^m \omega(A_iA_i^*) \geq 0 \ ,
\]
where we used condition 3 to obtain the first equality. In particular, setting
$A_i=\mathbf{1}$\ and $m=1$, this requires $\omega(Z) \in \R$, which
guarantees that $\left( \omega(Z)^2\right)^m \geq 0$\ for any $m$. The above
conditions therefore reduce to a smaller set of requirements
\[
\omega(A_iA_i^*) \geq 0 \ .
\]
We note that, unlike conditions 1 and 3, condition 2 is not linear: applying
it to linear combinations of basis elements may lead to additional independent
conditions on the values assigned by an almost-positive state.

Almost-positive states evolve in time $Z$\ using the flow generated by
$C=E+H$\ itself (which, as we noted earlier, preserves almost-positivity). Let
$\omega_{t} $\ be a one-parameter family of states along the flow generated by
$C$, with $t_0\in \R$\ being the initial value of the clock:
$\omega_{t_0} (Z) = t_0$. Then the clock evolves according to
\[
\frac{{\rm d}}{{\rm d} t} \omega_t (Z) = \frac{1}{i\hbar} \omega_t ([Z, C]) = 1
\]
so that $\omega_t(Z)=t$.

Since almost-positive states are completely characterized by the values they
assign to $Z$\ and the fashionables, the flow is defined by the way it affects
those values. However, it is useful to compute the time-evolution in $Z$\ for
an arbitrary element of $Z'$\ projected to fashionables. Using the fact that
$\alpha_t$\ is a $*$-algebra homomorphism and that it does not change the
expectation value in an almost-positive state, we have
\begin{eqnarray*}
&&\frac{{\rm d}}{{\rm d} t} \omega_t \left( \alpha_t \left(A_iZ^m\right)
   \right)\\
  &=&
  \frac{{\rm d}}{{\rm d} t} \omega_t \left( A_iZ^m \right)  \\
&=& \frac{1}{i\hbar} \omega_t ([A_iZ^m, E+H])
\\
&=& \omega_t \left( \frac{1}{i\hbar} [A_iZ^m, H]+  m A_iZ^{m-1}\right)
\\
&=& \omega_t \left( \frac{1}{i\hbar} [\alpha_t(A_iZ^m), \alpha_t ( H)] +
    \frac{{\rm d}}{{\rm d}t} \alpha_t (A_iZ^m) \right)  \ .  
\end{eqnarray*}
By linearity, this extends to an arbitrary $A\in Z'$
\begin{equation} \label{eq:ZEvolution}
\frac{{\rm d}}{{\rm d} t} \omega_t \left( \alpha_t \left(A\right) \right) = \omega_t
\left( \frac{1}{i\hbar} [\alpha_t(A), \alpha_t ( H)] + \frac{{\rm d}}{{\rm d}t} \alpha_t
  (A) \right)  \ .  
\end{equation}

With our specific choice of fashionables we have reproduced the usual form of
the quantum dynamical flow of equation~(\ref{eq:Hflow}). The fashionable
$\alpha_t ( H)$\ plays the role of the physical Hamiltonian associated with
the clock $(Z, \mathcal{F})$, generating evolution in $Z$\ directly on
$\mathcal{F}$, where any $A\in Z'$\ can be projected
using~(\ref{eq:FProjection}) and evolved using~(\ref{eq:ZEvolution}). Note
that if $H\in \mathcal{F}$, then $\alpha_t(H)=H$, however, in general, the
physical Hamiltonian is time-dependent.

\subsubsection{Fashionable ambiguities}\label{sec:Fashionables}

We conclude this subsection by focusing on the ambiguity associated with the
choice of fashionables that remains once a clock variable $Z$ is already
selected. In the case of a canonical clock discussed here, this ambiguity can
be linked to the freedom in selecting the conjugate momentum of the clock $E$
(by demanding that $[\mathcal{F}, E]=\{0\}$). Clearly, almost-positivity is
unaffected by the choice of fashionables. In addition, for any $A\in Z'$\ the
evolution of its expectation value relative to the clock $Z$ is given by
${\rm d} \omega_t (A)/{\rm d}t = -i\hbar^{-1} \omega_t ([A, C])$, which does
not rely on the choice of $\mathcal{F}$ either. In what way then does the
choice of fashionables matter? In order to reduce a completely constrained
system to an unconstrained system evolving in time, in addition to selecting
the measurement corresponding to time, one must also select a subset of
clock-compatible observables to be interpreted as measuring the state of ``the
rest of the system'' at any given time. If, for example, the expectation
values of all those measurements remain unchanged as the clock evolves, we
interpret ``the rest of the system'' as being static.

For a simple illustration of the freedom associated with choosing fashionables
for a given clock observable, consider the situation where $\mathcal{A}_S$\
corresponds to a one-component canonical system. The kinematical algebra
$\mathcal{A}\cong \mathcal{A}_T \otimes \mathcal{A}_S$\ is then generated by
four basic elements with $[Z, E]=i\hbar \mathbf{1} = [Q, P]$. This two-component
canonical algebra can just as well be generated using a different basis of
generators, for example $\{Z, \tilde{E}, Q, \tilde{P}\}$, where we define
$\tilde{E} = E+Q$, and $\tilde{P}=P-Z$. It is straightforward to check that
this basis generates the same algebra, that
$[Z, \tilde{E}]=i\hbar \mathbf{1} = [Q, \tilde{P}]$, and that all other
commutators between the generators vanish. Essentially, the new basis provides
an alternative factorization of $\mathcal{A}$\ into two canonical components,
where both factorizations are compatible with $Z$\ serving as the clock. For
the original generating basis the natural choice of fashionable algebra is
$\mathcal{F} = \C[Q, P]$\ (compatible with treating $E$\ as the generator of
time translation in $Z$), while for the second basis it is
$\tilde{\mathcal{F}} = \C[Q, \tilde{P}]$\ (compatible with treating
$\tilde{E}$\ as the generator of time translation in $Z$). It is
straightforward to convince oneself that the two fashionable algebras are not
the same: for example, $\tilde{P} \in \tilde{\mathcal{F}}$, but
$\tilde{P}=P-Z$\ is not an element of $\mathcal{F}$, since it cannot be
generated by polynomials in $P$ and $Q$ alone.

To see where the difference between these two choices of fashionable algebras
matters, consider the very simple case where $C=E=\tilde{E}-Q$. Reducing this
constraint using the clock $(Z, \mathcal{F})$\ results in a one-component
canonical system (generated by $Q$\ and $P$) that is static, since its
physical Hamiltonian is $H=C-E=0$. Performing reduction using
$(Z, \tilde{\mathcal{F}})$\ also leaves us with a one-component canonical
system (generated by $Q$\ and $\tilde{P}$), which, in this case, is not
static, since in this case the physical Hamiltonian
$\tilde{H}=C-\tilde{E}=-Q$\ does not vanish. Viewed as self-contained systems
the two reductions look different. At the same time, any element of $Z'$ can
be projected into either reduced system and its expectation value can be
evolved in time with identical end result.

It is instructive to see how this works out in our simple example. For
instance, $P\in Z'$ commutes with $C$: it is a physical observable, and
should, therefore, project to a constant of motion for any reduction of the
system. Using $(Z, \mathcal{F})$\ we have $\alpha_t(P) = P$, and
\begin{eqnarray*}
\frac{{\rm d}}{{\rm d} t} \omega_t \left( \alpha_t \left(P\right) \right) &=& \omega_t
\left( \frac{1}{i\hbar} [P, \alpha_t ( H)] + \frac{{\rm d}}{{\rm d}t} P
                                                                              \right)\\
                                                                              &=&
\omega_t \left( \frac{1}{i\hbar} [P, 0] \right) = 0 \ . 
\end{eqnarray*}
Using $(Z, \tilde{\mathcal{F}})$\ we have
$\tilde{\alpha}_t(P) = \tilde{\alpha}_t(\tilde{P}+Z) =\tilde{P}+t\mathbf{1}$,
and
\begin{eqnarray*}
\frac{{\rm d}}{{\rm d} t} \omega_t \left( \tilde{\alpha}_t \left(P\right) \right) &=&
\omega_t \left( \frac{1}{i\hbar} [\tilde{P}+t\mathbf{1}, \alpha_t (
                                                                                      \tilde{H})] + \frac{{\rm d}}{{\rm d}t} (\tilde{P}+t\mathbf{1}) \right)\\
  &=&  \omega_t
\left( \frac{1}{i\hbar} [\tilde{P}, -Q] +\mathbf{1}\right) = 0 \ . 
\end{eqnarray*}

The difference between a pair of reduced systems that use the same clock but a
different set of fashionables may be treated as superficial if the link to the
original constrained system is maintained. In addition, one particular choice
of generators may well be preferred on physical grounds of corresponding to a
natural set of measurements.

\section{Deparameterization by factorization}\label{sec:Factorization}

As we saw in section~\ref{sec:LinearConstraintAlgebra}, the conditions for
algebraic deparameterization are extremely restrictive, essentially requiring the constraint to have the form~(\ref{eq:ParamParticleC}). There is an important class of constraints that do not
have the simple form of the parameterized Newtonian particle but can,
nevertheless, be straightforwardly deparameterized using Hilbert space
methods. In particular, the motion a free relativistic particle on a Minkowski
space-time cast in the Hamiltonian form results in a constraint of the form
\[
C=E^2-H^2
\]
where $[H, E]=0=[H, Z]$. In this situation, the constraint explicitly
factorizes into two commuting factors $C=(E+H)(E-H)$. Either factor can play
the role of a constraint in its own right. In Hilbert space terms
$(E\pm H) | \psi \rangle =0$\ implies $C| \psi \rangle =0$. Conversely, the
Hilbert space solutions to the constraint are linear combinations of (generalized) solutions
to $(E\pm H)$. In this section we describe and apply the algebraic analog of
deparameterization of factorizable constraints first developed
in~\cite{AlgebraicTime}.

In our algebraic method, given a constraint $C$\ that does not satisfy the
definition of algebraic deparameterizability, we attempt to factorize it. A constraint $C$ is {\bf deparameterized by factorization} with respect
to an internal clock $(Z, \mathcal{F})$, if there are $N, C_H \in \mathcal{A}$, such that $C = N C_H$, where $C_H=C_H^*$\ is not a divisor of zero, has no inverse in $\mathcal{A}$, and is deparameterized by $(Z, \mathcal{F})$. The system with a factorized constraint is then reduced by deparameterizing $C_H$\ instead of $C$. We call $C_H$\ the \emph{factor constraint} and $N$\ the \emph{flow rate} of $C$\ with respect to $C_H$, since
$[Z, C]=i\hbar [N, Z] C_H + i\hbar N$, and hence for an almost-positive state relative to this deparameterization
\begin{equation}\label{eq:FlowRate}
\omega([Z, C])=i\hbar \omega(N) = \omega(N) \omega([Z, C_H]) \ .
\end{equation}

In order for this factorization procedure to make sense within our algebraic approach, we need to establish the equivalence between the solutions to the original constraint $C$\ and the factor constraint $C_H$. Recall that in section~\ref{sec:PhysicalStates} we defined physical states of $C$\ as orbits generated by the constraint flows on the constraint surface $\Gamma_{\rm phys} = \Gamma_C/\thicksim_C$. In the exceptional case where $N$\ is invertible in $\mathcal{A}$, we have $\mathcal{A}C=\mathcal{A}C_H$\ and  algebraically imposing the factor constraint is exactly equivalent to
imposing the original constraint. In general, however, since
$\mathcal{A}C = \mathcal{A}NC_H \subset \mathcal{A} C_H$, we have
$\Gamma_{C_H} \subset \Gamma_C$.  Solutions of the factor constraint $C_H$\ are also solutions of the original constraint $C$, but the converse is not necessarily true. Therefore, not all of the physical states of the original constraint $C$\ will be sampled by the deparameterization of the factor constraint $C_H$. This is not necessarily problematic: as we will see in the explicit example of section~\ref{sec:RelParticle}, the original constraint may have other factor constraints that sample other parts of its solution space. Moreover, we do not, in general, expect that all solutions to a factorizable constraint can be interpreted as evolving in time. A general Hamiltonian
constraint may possess several factorizations, relative to one or more clocks
where a given physical state may be sampled by some, but not other
factorizations, or by none at all.

The situation with gauge orbits is a bit more complicated: every gauge flow generator of the original constraint $AC=ANC_H$\ is also a gauge flow generator of the factor constraint, but, again, the converse is not
generally true. Let us look at the different gauge orbits $[\omega]_{C_H}$\ and $[\omega]_C$\ more closely. (This paragraph provides further details of a motivating discussion in our Introduction.)  Assuming factor $N$\ does not have an inverse within $\mathcal{A}$, $\mathcal{A}C$\ is a
proper subset of $\mathcal{A}C_{H}$, and hence $[\omega]_C \subset [\omega]_{C_H}$: the original orbits of $C$\ are contained within the larger orbits of $C_{H}$. As a result, some gauge flows generated by the factor constraint $C_{H}$\ are entirely new and can link distinct gauge orbits of the original constraint $C$. Therefore, a physical state with respect to $C_{H}$\ generally corresponds to a \emph{region} of the space of physical states with respect to the original constraint $C$. Furthermore, $C_{H}$, which is the driver of deparameterized evolution relative to the internal clock $Z$, is itself \emph{not} an element of $\mathcal{A}C$. This means that we cannot, in general, assume that $\omega \in \Gamma_{C_{H}}$\ and $S_{C_H}(\lambda) \omega$\ correspond to the same physical state relative to the original constraint $C$. Our construction in Section~4.1 of~\cite{AlgebraicTime} addresses this problem by developing conditions under which the gauge flows of the factor constraint $C_H$\ preserve the values assigned to the observable algebra of the original constraint $C$. We discuss it it in detail in the next section.

Before we move to a more detailed discussion of factorization, we note that the $*$-invariance of both the original and the factor constraints places a
general restriction on the flow rate. Combining $C=C^*$\ with $C_H=C_H^*$ gives $NC_H = C_HN^*$, or, equivalently,
\begin{equation} \label{eq:AdjRelation}
[N, C_H] = C_H (N^*-N) \ .
\end{equation}
If $N^*\neq N$, equation~(\ref{eq:FlowRate}) would generally lead to non-zero ${\rm Im}[\omega(Z)]$\ along the flow generated by the original constraint $C$. This would make it difficult to interpret deparameterization by factorization with respect to $Z$\ as a quantum version of ``re-scaling'' of the flow of $C$\ so that it is parameterized by $Z$. In what follows we will therefore often focus on factorized constraints with a \emph{real flow rate} $N^*=N$. According to the adjointness relation~(\ref{eq:AdjRelation}) such a flow rate will also necessarily be \emph{constant} $[N, C_H] = 0$.

\subsection{Observables of the factorized constraint}\label{sec:FactorObservables}

According to the discussion in section~\ref{sec:GaugeSections}, since the factor constraint $C_H$\ is deparameterized by the clock $(Z, \mathcal{F})$, we can use an almost-positive state in place of the entire corresponding gauge orbit $[\omega]_{C_H}$. Furthermore, the time evolution flow generated by $C_H$\ moves along the gauge orbits. This construction guarantees that the values of observables of $C_H$\ are fixed along the orbit, in particular making them constant along the time evolution generated by $C_H$. In the case of deparameterization by factorization, however, the true physical constraint is $C=NC_H$. As we have noted in the previous section, $[\omega]_C\subset[\omega]_{C_H}$\ so that an orbit of $C_H$\ will in general contain multiple orbits of the original constraint. This could lead to the orbit $[\omega]_{C_H}$\ containing states that assign different values to the observables of $C$, making an almost positive $\omega$\ a poor representative state and opening up the possibility that the values of these observables will change along the time-evolution flow of $C_H$. In this section we derive conditions that prevent this type of pathology.

Let $O_H$\ be in the observable algebra of the factor constraint so that
$[O_H, C_H]=0$. Then, relative to the original constraint,
\[
[O_H, C] = [O_H, NC_H] = N[O_H, C_H] + [O_H, N] C_H =  [O_H, N] C_H  \ .
\]
Due to the cancellation property of $C_H$, $O_H$\ is in the observable algebra of $C$\ precisely when $[O_H, N]=0$. However, even if $[O_H, N] \neq 0$, we\ get $\omega(A[O_H, C])=0$ for any solution of the factor constraint $\omega \in \Gamma_{C_H}$\ and any $A\in \mathcal{A}$. Effectively, $O_H$\ is an observable of the original constraint when we restrict to the states on the constraint surface of $C_H$. In particular, $O_H$\ has a unique value within each physical state of the original constraint, that also belongs to $\Gamma_{C_H}$. 

On the other hand, suppose $O$\ is an observable of the original
constraint. In order for its value to be preserved by all gauge flows
generated by $C_H$\ via equation~(\ref{eq:Flow}) starting from an initial
almost-positive state $\omega$, we need
\begin{equation}\label{eq:FactNecessary}
S_{AC_H}(\lambda) \omega([O, AC_H]) = S_{AC_H}(\lambda)  \omega(A[O, C_H]) = 0\ , \quad {\rm for\ all} \ \ A\in\mathcal{A} \ .
\end{equation}
However, $[O, C]=0$\ gives us $[O, NC_H] = [O, N] C_H + N[O, C_H]=0$. For an almost positive state $\omega \in \Gamma_{C_H}$, this gives us
\begin{equation}\label{eq:FactGiven}
S_{AC_H}(\lambda) \omega(AN[O, C_H])=0\ , \quad {\rm for\ all} \ \ A\in \mathcal{A} \ ,
\end{equation}
which is not sufficiently strong. In general, only the elements that commute with
both factors $N$\ and $C_H$\ will have gauge-independent values relative to
the physical states of both the original constraint $C$\ and those of the
factor constraint $C_{H}$. If one has a complete set of observables of $C$, as in the example of section~\ref{sec:RelParticle}, one can explicitly check condition~(\ref{eq:FactNecessary}). In the special case where $N\in Z'$, as in the class of constraints discussed in section~\ref{sec:LinearC}, we show that~(\ref{eq:FactNecessary}) holds as long as $N$\ is not a divisor of zero.

More generally, in~\cite{AlgebraicTime} we introduce an additional condition on states that makes condition~(\ref{eq:FactGiven}) imply condition~(\ref{eq:FactNecessary}), allowing us to interpret the physical states of $C_H$\ as states on the observable algebra of $C$: Left multiplication of
$A\in \mathcal{A}$ can be {\bf canceled in $\omega \in \Gamma$} if for any
$B \in \mathcal{A}$, $\omega(GAB) = 0$ for all $G\in \mathcal{A}$ implies
$\omega(GB) = 0$ for all $G\in \mathcal{A}$.  This condition is related to but
is not the same as requiring $A$\ not to be a divisor of zero. Indeed, if
$AB=0$\ for some $B\neq0$, then left multiplication of $A$\ cannot be canceled
in any state, since this implies $\omega(GAB)=0$\ for all $G$, while
$B\neq0$. While $A$\ not being a divisor of zero is necessary, it is not
sufficient for left cancellation, which imposes additional conditions on the
state. Suppose left multiplication of $N$
can be canceled in $\omega\in \Gamma_{C_H}$, so that for any $B\in \mathcal{A}$
\begin{equation} \label{eq:FactSufficient}
{\rm if} \ \ \omega(ANB) = 0 \ , \ \  {\rm for\ all\ } A\in\mathcal{A} \ , \ \  {\rm then}\ \ \omega(AB) = 0 \ \  {\rm for\ all\ } A\in\mathcal{A} \ .
\end{equation}
Then lemma~15 of~\cite{AlgebraicTime} demonstrates that, the values assigned to the elements that commute with the original constraint $C$\ are constant along the entire gauge orbit $[\omega]_{C_H}$\ generated by all of $\mathcal{A}C_H$.

We can somewhat relax the left cancellation condition in the case where the flow rate is {\em constant}, so that $[N, C_H]=0$. In Appendix~\ref{app:ConstantFlow Lemma}, where we appropriately specialize Lemma~15 of~\cite{AlgebraicTime} to the constant flow rate scenario. We show that, for a constant flow rate, the values of observables are preserved along orbits generated by $C_H$\ starting from some state $\omega \in \Gamma_{C_H}$\ as long as for any $B\in Z'$
\begin{equation} \label{eq:FactSufficient2}
{\rm if} \ \ \omega(ANB) = 0 \ , \ \  {\rm for\ all\ } A\in Z' \ , \ \  {\rm then}\ \ \omega(AB) = 0 \ \  {\rm for\ all\ } A\in Z' \ .
\end{equation}
The important relaxation in above is that left cancellation of
$N$, provided $[N, C_H]=0$, needs to be checked only for $A, B \in Z'$, rather
than $\mathcal{A}$.

Even in this relaxed formulation, however, given a state, it is difficult to explicitly determine whether the left cancellation condition holds. Fortunately, in the case of the real flow rate, we can use almost positivity to simplify it further. We recall from the discussion in section~\ref{sec:LinearCGauge} that for any $C_H$\ deparameterized by the clock $(Z, \mathcal{F})$ the algebra $\mathcal{A}$\ splits, so that for any $B\in Z'$\ left multiplication by the flow rate can be decomposed $NB=B_0 + B_1$\ where $B_0 \in Z'$\ and $B_1\in \mathcal{A}C_H$. Because these subsets are disjoint linear subspaces, the decomposition is linear, but does not preserve algebraic multiplication (except left multiplication by elements
of $Z'$). An almost-positive state assigns zero to elements of
$\mathcal{A}C_H$, so that $\omega(NB) = \omega(\hat{N}(B) )$, where
$\hat{N} \colon B \mapsto B_0$. In the simple case where $N \in Z'$\ we have $\hat{N}B = NB$. In general, $N=N_0 + \sum_{n=1}^M N_n C_H^n$\ for some integer $M$\ and $N_n\in Z'$. For convenience, we will denote repeated commutator with the factor constraint as ${\rm ad}_{C_H}^n A$, where ${\rm ad}_{C_H}^0A = A$\ and ${\rm ad}_{C_H}^{n+1} A=[{\rm ad}_{C_H}^n A, C_H]$. Permuting factors of $C_H$\ to the right one-by one (recall from section~\ref{sec:LinearCGauge} that ${\rm ad}_{C_H}$\ preserves $Z'$) we note that
\[
C_H^n B =  (-1)^n {\rm ad}_{C_H}^n B + G C_H \ ,
\]
where $G\in \mathcal{A}$\ is a combination of $B$, $C_H$\ and their commutators. Thus
\begin{equation} \label{eq:N-hat}
\hat{N} B = N_0 B + \sum_{n=1}^M N_n (-1)^n {\rm ad}_{C_H}^n B \ .
\end{equation}

The left cancellation condition for a real flow rate~(\ref{eq:FactSufficient2}), evaluated in an almost-positive state then has the general property that
$\omega\left(A\hat{N}(B) \right) = 0$\ for all $A\in Z'$
implies $\omega\left(AB \right) = 0$\ for all $A\in Z'$. Because
$\omega$\ is positive on $Z'$,
$\omega\left(A\hat{N}(B) \right) = 0$\ for all $A\in Z'$\
if and only if
$\omega\left(\hat{N} (B)^* \hat{N} (B) \right) =
0$.\footnote{Note that, in general,
  $\hat{N} (B)^* \neq \hat{N} (B^*)$.}  The necessity is
trivial, sufficiency follows from the Schwarz-type inequality
\[
\left|\omega\left(A\hat{N} (B) \right) \right|^2 \leq \omega \left(AA^*\right)
\omega\left(\hat{N} (B)^* \hat{N} (B)  \right) \ . 
\]
A more compact version of the cancellation restriction is therefore to only
consider almost-positive states such that
$\omega\left(\hat{N} (B)^* \hat{N} (B) \right) = 0$\ implies
$\omega(B^*B)=0$\ for any $B\in Z'$. A positive linear functional
(PLF) on $Z'$\ can be used to construct a pre-Hilbert space
representation $\Lambda(Z')$; if $\omega((AB)^*AB)=0$\ in this PLF,
then $\Lambda(A)$\ has zero in the discrete part of the spectrum, so one way
to look for appropriate states within a representation $\Lambda$\ of
$Z'$\ is to study the spectra of $\Lambda( \hat{N} (B) )$.

We see that deparameterization via factorization successfully casts a quantum system with a Hamiltonian constraint as an unconstrained dynamical system, with one important caveat. This is only possible for physical states that contain some almost--positive states relative to deparameterization of $C_H$, in which condition~(\ref{eq:FactNecessary}) can be verified either directly or via left cancellation of the action of the flow rate factor $N$.

\subsection{Linear factorizable constraint}
\label{sec:LinearC}

Let us assume that $C=NC_H$\ is deparameterizable with respect to
$(Z, \mathcal{F})$\ by factorization, and consider the simplest generalization
of the directly-deparameterizable situation, namely that
$[Z, C] \neq i\hbar \mathbf{1}$, but $[Z, [Z, C]]=0$. From the conditions
placed on $C$, $C_H$\ and $Z$, it then follows that such a constraint can be
written in the form
\[
C=B_0 + B_1 C_H \ ,
\]
for some $B_0, B_1 \in Z'$. The constraint is therefore \emph{linear} in
momentum conjugate to the clock variable $Z$ (in this case represented by
$C_H$). This can be used to demonstrate that $[N, Z]=0$, $N^*=N$, and
$[N, C_H]=0$, so that the flow rate of a linear factorizable constraint is automatically real (and therefore also constant). Details of the argument can be found in Appendix~\ref{app:LinearC}. 

Do the gauge flows of the factor constraint $C_H$\ preserve the values that states assign to observables of the original constraint $C$? In this simple scenario, we can ensure this in two distinct ways. First, any $O\in \mathcal{A}_{\rm obs}$\ can be written as $O=O_0+ O_1C_H$\ for some $O_0 \in Z'$\ and $O_1 \in \mathcal{A}$. For $O\in \mathcal{A}_{\rm obs}$\ we then have
\[
0 = [O, C] = N[O_0, C_H]+\left( N[O_1, C_H] C_H + [O,N] C_H \right) \ .
\]  
Since $N\in Z'$\ for a linear constraint and $[O_0, C_H]\in Z'$\ for any $O_0\in Z'$, the first term in the final expression above is in $Z'$, while the rest of the expression is clearly in $\mathcal{A}C_H$. Since the two subalgebras are disjoint by the requirement of deparameterization, the two parts of the final expression must vanish separately; in particular
\[
N[O_0, C_H] = 0 \ .
\]
Provided that $N$\ is not a divisor of zero within $Z'$, this implies $[O_0, C_H]=0$, so that in any state $\omega \in \Gamma_{C_H}$
\[
S_{AC_H}(\lambda) \omega(A[O, C_H]) = S_{AC_H}(\lambda) \omega (B[O_1, C_H]C_H) = 0 \ ,
\]
for any $A\in \mathcal{A}$\ as required by condition~(\ref{eq:FactNecessary}).

We can also apply the left cancellation condition. We note that the flow rate here is real and that $N\in Z'$. The left cancellation condition therefore simplifies to: for any $A\in Z'$\ if $\omega \left( (NA)^* (NA) \right)=0$\ then $\omega(A^*A)=0$. Formulated in this way, the left cancellation property is related to the spectrum of $N$\ in a Gelfand-Naimark-Segal representation containing $\omega$, and one might intuit that, if $N$\ is
positive definite, there should not be any almost-positive states that violate
the left cancellation condition. Indeed there are ways to ensure this. For
example, let $N = BB^*+a \mathbf{1}$\ for some $B \in Z'$\ and some real number $a>0$, then
\begin{eqnarray*}
\omega \left( (NA)^* (NA) \right) &=& \omega \left[ \left( \left( BB^*+a
                                      \mathbf{1} \right) A\right)^*
                                      \left( \left( BB^*+a \mathbf{1} \right)
                                      A \right) \right] \\  
&=& \omega \left[ \left( BB^*A \right)^* \left(BB^*A \right) \right] + 2 a\,
    \omega \left[ \left( B^*A \right)^* \left(B^*A \right) \right] + a^2
    \omega \left[ A^*A \right] \ . 
\end{eqnarray*}
Each of the terms in the final expression is proportional to $\omega(D^*D)$, for a $D\in Z'$. For an almost-positive $\omega$\ each term is non-negative, therefore if $\omega \left( (NA)^* (NA) \right) = 0$, each term must vanish separately. Thus, in particular, $\omega \left( (NA)^* (NA) \right) = 0$\ here would automatically imply $\omega(A^*A) = 0$. In this specific case, the left action of $N$\ within $Z'$\ would be canceled in any almost-positive state.

\subsection{Factorizable constraints with a canonical clock}
\label{sec:CanonicalFactor}

Here we make the same assumptions about the kinematical algebra as in
Section~\ref{sec:LinearConstraintAlgebra}. A constraint that is
deparameterizable by factorization must then have the following general form
\[
C=NC_H=N(E+H) \ .
\]
Where $N \in \mathcal{A}$, $H=H^* \in Z'$\ and $E$\ is canonically conjugate
to $Z$. We assume that the fashionable algebra commutes with $E$, so that
$Z'$\ is algebraically generated by $\{Z\}\cup\mathcal{F}$\ and $\mathcal{A}$\
is algebraically generated by $\{E\} \cup Z'$ (as well as by $\{C_H\}\cup Z'$\
as required by deparameterization). It is convenient to classify such constraints by their polynomial order in
$E$. For a constraint linear in $E$, we immediately have $[Z, [Z, C]]=0$\ and
the results of the previous section apply: $N\in Z'$, $N^*=N$\ and $[N, C_H]=0$,
so that $[N, E] = -[N, H]$. If, in addition, $N$\ is not a divisor of zero within $Z'$, the deparameterization by factorization preserves the observables of the original constraint $C$.

To see just how restrictive the adjointness conditions are, let us consider the ``nice--looking'' linear constraint of the form
\begin{equation}\label{eq:LinearExC}
C=\frac{1}{2 } (B_1E + EB_1) + B_0 \ ,
\end{equation}
with $*$-invariant $B_0$ and $B_1$ such that $[B_1,E]\not=0$ and $B_1$ is invertible. Noting that  the ``unfactorized'' constraint must have the form  $C=NE+NH$, we reorder and factorize
\begin{eqnarray*}
  C&=&B_1E+B_0-\frac{1}{2}[B_1,E]\nonumber\\
  &=& B_1\left(E+\frac{1}{2}(B_1^{-1}B_0+B_0B_1^{-1})+
      \frac{1}{2}[B_1^{-1},B_0]- \frac{1}{2}B_1^{-1}[B_1,E]\right) \ .
\end{eqnarray*}
So that $N=B_1$, which is $*$--invariant, while
\begin{equation}
  C_H=E+\frac{1}{2}(B_1^{-1}B_0+B_0B_1^{-1})+
      \frac{1}{2}[B_1^{-1},B_0]- \frac{1}{2}B_1^{-1}[B_1,E] \ .
\end{equation}
The first two terms in the expression fo $C_H$\ are always $*$--invariant, however, unless additional conditions are imposed, the sum of the last two terms is not. After a bit of algebra one finds that $C_H=C_H^*$\ is equivalent to
\begin{equation}\label{eq:LinearExCadj}
[B_0, B_1] = \frac{1}{2} \left( B_1[B_1, E] + [B_1, E]B_1 \right) \ .
\end{equation}
We conclude that the constraint of equation~(\ref{eq:LinearExC}) cannot be deparameterized relative to clock $Z$, unless the additional condition of equation~(\ref{eq:LinearExCadj}) is satisfied. (Note that, because $C=C^*$\ here, $C_H=C_H^*$\ and $N=N^*$\ together automatically imply $[N, C_H]=0$, so the latter condition does not generate additional restrictions.) In this example, there is nothing preventing us from defining almost-positive states that solve $C_H$\ and are positive on $Z'$\ and fashionables. However, the gauge flow of a non $*$-invariant $C_H$\ implies that evolution of fashionables would not preserve their reality and cannot be interpreted as unitary time-evolution.

A more interesting scenario in the context of quantum cosmology is the
situation where $C$\ is quadratic in $E$. Assuming $C$\ can be factorized,
\begin{equation}\label{CFac}
C=(N_1E+N_0)(E+H) \ .
\end{equation}
for some $N_0, N_1, H \in Z'$, with $H^*=H$. Here as well, exact
factorizability places strong restrictions on the possible form of the
constraint, such as the factor ordering chosen for the original quadratic
expression. Our algebraic conditions on deparameterization allow us to derive
additional constraints on the terms in (\ref{CFac}), following the discussion
in Section~4.2 of~\cite{AlgebraicTime}. To this end, we apply algebraic
deparameterization to (\ref{CFac}) such that $C_H=E+H$ is the linearized
constraint with flow rate $N=N_1E+N_0$. The adjointness conditions $C=C^*$, $Z=Z^*$\ by themselves give 
\[
\left( \frac{1}{i\hbar} \left[ Z, \frac{1}{i\hbar} [Z, C] \right] \right) =
\left( \frac{1}{i\hbar} \left[ Z, \frac{1}{i\hbar} [Z, C] \right] \right)^* \ ,
\]
which immediately implies $N_1=N_1^*$. Unlike the linear case, here the adjointness conditions on their own do not force the flow rate to be real (or equivalently to be constant). 

Restricting to the cases where $N$ is required to be
real, $N^*=EN_1+N_0^*=N=N_1E+N_0$, we obtain
\begin{equation} \label{N0}
  N_0^*=N_0+[N_1,E]\,.
\end{equation}
According to~(\ref{eq:AdjRelation}), a real flow rate is also constant, $[N,C_H]=0$, which yields
\[
  0=([N_1,E]+[N_1,H])E+N_1[E,H]+[N_0,E]+[N_0,H]\,.
\]
The parenthesis as well as the last three terms of this equation are elements
of $Z'$. Taking a commutator of the whole equation with $Z$ therefore implies
\begin{equation} \label{N1EH}
  [N_1,E]+[N_1,H]=0\,.
\end{equation}
The remaining terms then require
\begin{equation}\label{N1N0}
  N_1[E,H]+[N_0,E]+[N_0,H]=0\,.
\end{equation}
The general factorized constraint (\ref{CFac}), when multiplied out, takes the
complicated form
\begin{equation}
  C=N_1E^2+(N_0+N_1H)E+N_1[E,H]+N_0H \ .
\end{equation}
The coefficients of powers of $E$\ are subject to $N_1=N_1^*$\ and, in the case of a real flow rate, also to equations~(\ref{N0}) through~(\ref{N1N0}). Can these conditions help us determine whether a general constraint that is a quadratic polynomial in $E$\ is deparameterizable by factorization? 

We are able to answer this question in the affirmative if 
$N_1=\mathbf{1}$. The flow rate then equals $N=E+N_0$ and (\ref{N0}) implies
that $N_0$ is $*$-invariant. Condition (\ref{N1EH}) is immediately
satisfied, while (\ref{N1N0}) reads
\begin{equation}\label{HE}
  [H,E]=[N_0,E]+[N_0,H]\,.
\end{equation}
Now the constraint has the simpler form
\begin{equation} \label{eq:CwithN1=1}
C=E^2+A_1E+A_0 \ ,
\end{equation}
with $A_1=N_0+H=A_1^*\in Z'$ and
$A_0=[E,H]+N_0H\in Z'$. We can perform factorization in two steps. First we complete the square with $E$
\begin{equation}\label{eq:CompleteSquare}
  C=\left(E+\frac{1}{2}A_1\right)^2-
  \left(\frac{1}{4}A_1^2-\frac{1}{2}[A_1,E]-A_0\right)= \tilde{E}^2-h
\end{equation}
where $\tilde{E}=E+\frac{1}{2}A_1$ and
\begin{equation}\label{eq:h}
  h = \frac{1}{4}A_1^2-\frac{1}{2}[A_1,E]-A_0 \ .
\end{equation}
Here $\tilde{E}$ is $*$-invariant because $A_1^*=A_1$, and it has the canonical commutator with $Z$, $[Z,\tilde{E}]=i\hbar\mathbf{1}$, because $A_1\in Z'$. Clearly $h\in Z'$\ because $A_1$\ and $A_2$\ are in $Z'$. Using the definitions of  $A_0$ and $A_1$\ we have
\begin{eqnarray*}
  h&=&\frac{1}{4}\left( N_0 + H\right) ^2-\frac{1}{2}[N_0,E]+\frac{1}{2}[H,E]-N_0H
\\
&=& \frac{1}{4} (N_0^2 + H^2 + N_0H + HN_0)- \frac{1}{2}(N_0H+HN_0) 
\\
&=& \left( \frac{1}{2} (N_0-H) \right)^2 \ ,
\end{eqnarray*}
where we also used~(\ref{HE}) to obtain the second equality. Writing $h$\ in this way we immediately see that $h^*=h$, it has a square root $\sqrt{h} = \frac{1}{2} (N_0-H)$, and condition~(\ref{HE}) is equivalent to $[\sqrt{h},\tilde{E}]=0$. Clearly, when this condition is satisfied the difference of squares in equation~(\ref{eq:CompleteSquare}) can be factorized as
\[
C=(\tilde{E}-\sqrt{h}) (\tilde{E} + \sqrt{h}) = \left( E + \left( \frac{1}{2}A_1-\sqrt{h} \right) \right) \left(E + \left( \frac{1}{2}A_1+\sqrt{h}\right) \right) \ .
\]
What we have shown is that a quadratic constraint of the form~(\ref{eq:CwithN1=1}) is deparameterizable by factorization with respect to $Z$\ precisely when: $A_0, A_1 \in Z'$; $A_1=A_1^*$; $h$\ defined by~(\ref{eq:h}) is $*$-invariant and has a square root $\sqrt{h}$\ such that $[\sqrt{h},E+\frac{1}{2}A_1]=0$. Since the two factors commute and have canonical commutator with $Z$, in this case either factor can be moved to the right to play the role of $C_H$. We have not one, but two deparameterizations of $C$\ with respect to $Z$. We study the consequences of this more closely through a concrete example in the next section.

\subsection{Quadratic example: ``slow'' relativistic particle}
\label{sec:RelParticle}

The Hamiltonian constraint for the free particle (rest mass $\mu>0$) in
Minkowski spacetime relative to the standard coordinates is identical to the
relativistic energy-momentum relation
\[
E^2-p^2-\mu^2=0 \ ,
\]
where we have set the speed of light $c=1$\ and $p^2 = p_x^2+p_y^2+p_z^2$\
relative to Cartesian coordinates on space.  In order to work with a simple
polynomial algebra and avoid having to define general square-root elements, we
will assume that the particle is ``slow'', i.e. $p\ll \mu$. We write
$E^2-p^2-\mu^2 = E^2-(\frac{1}{2}p^2/\mu
+\mu)^2+p^2(\frac{1}{2}p/\mu)^2$. Dropping the last term the approximate
Hamiltonian constraint is
\begin{equation}\label{CRel}
C=E^2-\left(\frac{p^2}{2\mu} +\mu\right)^2 \ .
\end{equation}

We will further simplify matters by assuming that there is only one spatial
dimension, writing $p=p_x$, and $q=x$. As our quantum kinematical algebra
$\mathcal{A}$\ we will use the algebra of polynomials with complex
coefficients generated by two canonical pairs of variables with the
non-zero commutators having canonical form,
$[q, p] = [Z, E]=i\hbar \mathbf{1}$, where each generator is $*$-invariant. We
note that this algebra can be represented as operators on the space of
Schwartz-type wave functions on $\mathbb{R}^2$\ with the usual square-integral
inner product.

This constraint has the form of a difference of commuting squares we saw in equation~(\ref{eq:CompleteSquare}), with $\tilde{E}=E$\ and $\sqrt{h}=(\frac{1}{2}p^2/\mu +\mu\mathbf{1})$, which factorizes as
\[
C=C_{+}C_{-} = C_{-}C_{+}
\]
where $C_{\pm} = E \pm (\frac{1}{2}p^2/\mu +\mu\mathbf{1})$. The factors
commute, neither is a divisor of zero or has an inverse in $\mathcal{A}$, and
we also have $C_{\pm}^* = C_{\pm}$.\footnote{In the algebra of polynomials the
  only divisor of zero is $0$\ and only elements $\alpha\mathbf{1}$, with
  $\alpha\in \mathbb{C}$ are invertible.} Either factor can be used to define
a linearization of the constraint: If $C_{+}$\ plays the role of $C_H$, then
$C_{-}$\ plays the role of $N$\ and vice versa.

Each factor has the form of a parameterized Newtonian particle and can be
deparameterized by $Z$\ as the clock. To see this we note that $\mathcal{A}$\
has a basis of specially ordered monomials $q^kp^lZ^mE^n$, for integer
$k, l, m, n$, analogous to our example in Section~\ref{sec:NonHamCExample}. As
in Section~\ref{sec:LinearConstraintAlgebra}, the commutant $Z'$\ has a linear
basis consisting of monomials $q^kp^lZ^m$, restricting the basis of
$\mathcal{A}$ to those elements with $n=0$. The ideal $(Z-t\mathbf{1})Z'$\
defines cosets on $Z'$ given by
\[
\left[  q^kp^l  Z^m \right] = \left[  q^kp^l  t^m \right] = t^m \left[  q^kp^l  \right] \ .
\]
The collection of cosets $\left\{ \left[ q^kp^l \right] \right\}$, therefore
provides a linear basis on the quotient algebra $Z'/(Z-t\mathbf{1})Z'$, on
which the canonical
projection $\pi_t\colon Z' \rightarrow Z'/(Z-t\mathbf{1})Z'$  acts by
\[
\pi_t \left( q^kp^l  Z^m \right) = t^m \left[  q^kp^l  \right] \,.
\]
The linear span of the monomials $q^kp^l\in Z'$\ is the natural choice for the fashionable
algebra ${\cal F}$\ corresponding to treating $E$\ as the time translation generator, so that $[E, \mathcal{F}]=0$ (see discussion in section~\ref{sec:Fashionables}). The clock $(Z, \mathcal{F})$ deparameterizes each factor $C_{\pm}$\ as
described in Section~\ref{sec:AlmostPositive}.

Let us explicitly characterize the algebraic restrictions placed on
kinematical states by this deparameterization. An almost-positive state defined in section~\ref{sec:LinearCGauge} is a solution of $C_{\pm}$. Using the same factorization as above
$\omega_{\pm} (AC_{\pm}) = 0$\ yields a condition for the value assigned to
basis monomials
\begin{eqnarray*}
\omega_{\pm} \left( q^kp^lZ^mE^n \right) &=& (\mp 1)^n \omega_{\pm} \left(
                                             q^kp^lZ^m\left( \frac{p^2}{2\mu}
                                             +\mu\mathbf{1} \right)^n \right) \\ 
&=&   (\mp 1)^n \sum_{j=0}^{n} \left( \frac{1}{2\mu} \right)^j \mu^{n-j}
    \omega_{\pm} \left( q^kp^{l+2j}Z^m \right) \ . 
\end{eqnarray*}
As anticipated by the discussion in Section~\ref{sec:AlmostPositive}, the
above relation can be interpreted as placing no restrictions on the values
assigned by states to elements of $Z'$ (basis elements with $n=0$), which can
then be used to completely determine the values assigned to the rest of
$\mathcal{A}$. In addition, an almost positive state parameterizes $Z$, leading
to restrictions on values assigned to $Z'$
\[
\omega_{\pm} \left( q^kp^lZ^m \right) = \omega_{\pm} \left( Z \right)^m
\omega_{\pm} \left( q^kp^l \right) \ .  
\]
We can freely specify the values assigned by the state to the clock
$\omega_{\pm} \left( Z\right)$\ and to elements of the fashionable algebra; after that the values assigned to the rest of $Z'$\ are
completely fixed by the parameterization condition. Finally, positivity on
$Z'$\ is satisfied if, in addition,
$\omega_{\pm} \left( Z \right) \in \mathbb{R}$\ and $\omega_{\pm}$\ is
positive on $\mathcal{F}$, which captures the degrees of freedom of the deparameterized system. They evolve relative to clock $Z$\ along the flow generated by $C_{\pm}=E \pm \sqrt{h}$. Since $\sqrt{h}=(\frac{1}{2}p^2/\mu +\mu\mathbf{1})$\ is a fashionable and commutes with $E$, following the same reasoning as in section~\ref{sec:AlmostPositive} leading up to equation~(\ref{eq:ZEvolution}), deparameterized time evolution of values of $A\in\mathcal{F}$\ is generated via
\[
\frac{{\rm d}}{{\rm d} t} \omega_{t;\, \pm} \left( A \right) = \frac{1}{i\hbar}\omega_{t;\, \pm} \left(
 \left[ A, \pm \left(\frac{1}{2\mu}p^2 +\mu\mathbf{1}\right) \right] \right) \ .  
\]

Now that we understand how to algebraically deparameterize either factor
$C_{\pm}$\ with respect to $Z$, let us discuss how this deparameterization
relates to the algebraic solution of the original constraint $C$, as defined
in Section~\ref{sec:PhysicalStates}. Let us begin by understanding the
relation of the constraint surfaces $\Gamma_{C_{\pm}}$\ defined by the two
factors to the constraint surface $\Gamma_C$\ corresponding to the original
constraint.  Since both $\omega(AC_{+})=0$\ and $\omega(AC_{-})=0$\ also imply
$\omega(AC)=0$, every solution of $C_{\pm}$\ is also a solution of
$C$. Therefore both constraint surfaces $\Gamma_{C_{\pm}}$ are entirely contained
within the constraint surface $\Gamma_C$. Furthermore, normalized combinations
of states from $\Gamma_{C_+}$\ and $\Gamma_{C_-}$\ also give us solutions to
$C$. In particular, if $a_+, a_- \in \mathbb{C}$, where $a_++a_-=1$, and if we
have two states $\omega_+ \in \Gamma_{C_+}$ and $\omega_- \in \Gamma_{C_-}$,
then $\omega =a_+\omega_+ + a_-\omega_-$\ is an element of $\Gamma_C$.

For any constraint of the type $C=C_+C_-=(\tilde{E} + \sqrt{h}) (\tilde{E} - \sqrt{h})$ (discussed in section~\ref{sec:CanonicalFactor}), including the current
example, the two surfaces $\Gamma_{C_{\pm}}$\ are not disjoint. A solution to
both constraint factors must satisfy $\omega(AC_+) = 0$\ and
$\omega(AC_-) = 0$\ for any $A \in \mathcal{A}$. These conditions are
entirely equivalent to requiring that both $\omega(A\tilde{E})=0$\ and
$\omega(A\sqrt{h}) = 0$\ for all $A \in \mathcal{A}$, since
\begin{eqnarray*}
\omega(A\tilde{E}) &=& \omega \left(A\cdot\frac{1}{2} (C_+ +C_-) \right) = \frac{1}{2} \left(
  \omega(AC_+) + \omega(AC_-) \right) = 0 \ , \\
\omega(A\sqrt{h}) &=& \omega \left(A\cdot\frac{1}{2} (C_+ -C_-) \right) = \frac{1}{2}
               \left( \omega(AC_+) - \omega(AC_-) \right) = 0 \ . 
\end{eqnarray*}
Conversely, $\omega(A\tilde{E})=0$\ and $\omega(A\sqrt{h}) = 0$\ immediately imply both
$\omega(AC_+) = 0$\ and $\omega(AC_-) = 0$. Now, the only restriction on the
values assigned by a general state $\omega \in \Gamma$\ is normalization
$\omega(\mathbf{1})=1$. So, it is possible to satisfy both $\omega(A\tilde{E})=0$\ and
$\omega(A\sqrt{h}) = 0$\ for all $A$, \emph{unless} $A\tilde{E} + B\sqrt{h} = \mathbf{1}$\ for some
$A, B \in \mathcal{A}$. No such $A$\ and $B$\ exist within $\mathcal{A}$\ in
our example, hence the intersection $\Gamma_{C_+}\cap \Gamma_{C_-}$\ is
non-empty.

However if we consider only almost-positive states, there are additional
restrictions. In the present example, since $p=p^*\in Z'$,
\[
\omega(\sqrt{h}) = \frac{1}{2\mu}\omega\left( pp^*\right) + \mu \geq  \mu > 0 \ ,
\]
which means $\omega(\sqrt{h}) =0$\ cannot be satisfied by an almost--positive
state. Hence the sets of almost--positive states with respect to internal
clock $Z$\ defined by the two constraint factors are, in this case, completely
disjoint. Furthermore, because $[C_+, C_-]=0$, the value
$\omega_+(C_-)=\omega(C_+-2\sqrt{h})=-2\omega(\sqrt{h})<-2\mu \neq 0$\ is preserved along
the entire orbit generated by $\mathcal{A}C_+$, therefore
$[\omega_+]_{C_+} \cap \Gamma_{C_-} = \emptyset$. By a symmetric argument
$[\omega_-]_{C_-} \cap \Gamma_{C_+} = \emptyset$. The constraint of our
example possesses two factorizations deparameterizable by the same clock $Z$,
which sample distinct physical states of the original constraint (note that $[\omega]_C \subset [\omega]_{C_{\pm}}$\ as discussed in detail below).

Let us  consider the gauge orbits $[\omega]_{C_{\pm}}$\ generated by a factor constraint more closely. As we have pointed out in section~\ref{sec:FactorObservables}, $[\omega]_C \subset [\omega]_{C_{\pm}}$\ and we generally need additional conditions (see sections~\ref{sec:FactorObservables} and~\ref{sec:LinearC}) to ensure that the observables of the original constraint $C$\ are preserved along the gauge flows associated with a factor constraint $C_{\pm}$. In this particular case, it is straightforward to check this explicitly by finding a complete set of observables. The complete classical solution of a two--component system with a single constraint (that could then be quantized), should result in $2\times2-1=3$ independent observables that are constant along the flow generated by the constraint via the Poisson bracket. (One further degree of freedom would then be eliminated by the constraint condition, leaving us an one--component unconstrained system.) Clearly $[E, C]=[p, C]=0$\ and it is not very difficult to find the third independent observable $O=\frac{1}{\mu}Zp\sqrt{h}+qE$. It is immediately obvious that $E$\ and $p$\ also commute with the factor constraints $C_{\pm}$, and a brief calculation yields
\begin{eqnarray*}
[O, C_\pm] &=& \left[ \frac{1}{\mu} Zp\sqrt{h} + qE, E\pm\sqrt{h} \right]
\\
&=& \frac{1}{\mu} p\sqrt{h} [Z, E] \pm [q,\sqrt{h}] E
\\
&=& \frac{1}{\mu} i\hbar p\sqrt{h} \pm \frac{1}{\mu} i\hbar pE = \frac{i\hbar}{\mu}  p C_\pm \ .
\end{eqnarray*}
Therefore, for any $\omega \in \Gamma_{C_\pm}$\ all states within the orbit $[\omega]_{C_{\pm}}$\ assign identical values to the observable algebra $\mathcal{A}_{\rm obs}$\ associated with the original constraint $C$, here generated by $E$, $p$\ and $O$.

\section{Methods for approximate deparameterization} \label{sec:Approximations}

As we have seen in sections~\ref{sec:GaugeSections} and~\ref{sec:Factorization}, given an internal clock, only a select few of the constraints that satisfy the conditions laid out in section~\ref{sec:PhysicalStates} can be deparameterized by it exactly---either directly or by factorization. This should not be surprising: even in the classical description of totally constrained systems a given internal clock, in general, is only locally (and temporarily) valid where the Poisson bracket $\{Z, C\}$\ is non-vanishing. In the quantum case, the states are difficult to localize, especially in the kinematical setting where nothing restricts gauge orbits to be ``local'' and physical states can only be truly localized on the values they assign to the Dirac observables of the system. It is therefore reasonable to expect that quantum deparameterization will, in general, only hold \emph{approximately}, and only on states that satisfy additional localization conditions. In this section we discuss several state-based strategies for approximately deparameterizing a given constraint. Our objective here is to show that our algebraic approach to deparameterization is well-suited for development of approximation techniques, some of which will be briefly explored, leaving their detailed study for another time.

\subsection{Approximate factorization}\label{sec:ApproxFactorization}

Consider a pair of $*$--invariant elements $N, C_H \in \mathcal{A}$, where $C_H$\ has all the properties of a constraint that is exactly deparameterizable by some clock $(Z, \mathcal{F})$. For $[N, C_H] \neq 0$, the product $NC_H$\ is not $*$--invariant and so could not serve as a constraint. Consider instead the constraint
\[
C=NC_H-\frac{1}{2}[N, C_H] \ .
\]
It is straightforward to check that $C^*=C$, however $C_H$\ is not exactly a factor constraint of $C$, but it would be approximately a factor of $C$\ if $[N, C_H]$\ is ``small''. Since $[N, C_H]$\ is just some element of the kinematical algebra $\mathcal{A}$\ its value in a state is a priori not restricted by anything other than normalization and constraint conditions. Suppose $\omega$\ is a solution of $C_H$, under what conditions does it also solve $C$? Since
\[
\omega(AC) = \omega(ANC_H-\frac{1}{2}A[N, C_H]) = \frac{1}{2}\omega(A[N, C_H]) \ ,
\]
we need $\omega(A[N, C_H])=0$\ for all $A\in \mathcal{A}$, imposed on solutions to $C_H$\ in order for them to also solve $C$. Unfortunately, unless $[C_H, [N, C_H]]=0$, this leads to further conditions or inconsistencies, since we need
\[
\omega\left( [C_H, [N, C_H]] \right) = \omega\left( C_H [N, C_H] \right) +\omega\left( [N, C_H] C_H \right) = 0 \ .
\]
For example, suppose $N= a\mathbf{1}+bZ^2$, where $a, b\in \mathbb{R}$, we have $\omega\left( [C_H, [N, C_H]] \right) = 2\hbar^2b$, so that $C_H$\ and $C=\left( a\mathbf{1}+bZ^2\right)C_H-i\hbar b Z$\ do not share any solutions at all. Nevertheless, almost positive states of $C_H$\ for which
\[
\omega( A[N, C_H]) = 2i\hbar b\, \omega(AZ) = 2i\hbar \left( \omega(Z) \omega (A) + \omega([A, Z]) \right)
\]
is \emph{small} for all $A\in \mathcal{A}$\ are also approximate solutions to the constraint $C$. 

What general form can this ``smallness'' condition take? For a general flow rate $N=N_0+N_1 C_H$, where $N_0\in Z'$\ and $N_1 \in \mathcal{A}$, so that  in an almost positive state $\omega \left(A [N, C_H] \right) = \omega \left( A [N_0, C_H] \right)$. Further, for an arbitrary $A \in \mathcal{A}$\ there are $B_n \in Z'$\ and an integer $M$, such that $A = B_0 + \sum_{n=1}^M B_n C_H^n$. So that
\begin{equation}\label{eq:ApproxFactor}
\omega \left( A [N, C_H] \right) =  \sum_{n=0}^M \omega \left( B_n C_H^n [N_0, C_H] \right) 
=\sum_{n=0}^M \omega \left( B_n\, {\rm ad}_{C_H}^{n+1} (N_0) \right) \ ,
\end{equation}
where we have commuted every factor of $C_H$\ all the way to the right and used the fact that $\omega$\ solves $C_H$. We note that ${\rm ad}_{C_H}^n (N_0)$\ (using notation introduced in section~\ref{sec:FactorObservables} leading up to equation~(\ref{eq:N-hat})) is an element of $Z'$\ for any $n$. By positivity of $\omega$\ on $Z'$\ we therefore have $\omega \left( \left( {\rm ad}_{C_H}^n (N_0) \right)^* {\rm ad}_{C_H}^n (N_0) \right) \geq 0$. Let us restrict to almost positive states for which these values are suppressed by some small quantity $\epsilon$, formally 
\begin{equation}\label{eq:ApproxFactCond}
\sqrt{\omega \left( \left( {\rm ad}_{C_H}^n (N_0) \right)^* {\rm ad}_{C_H}^n (N_0) \right)} \propto \hbar^n \epsilon \ ,
\end{equation}
where we have included factors of $\hbar$\ to keep track of the number of commutators taken. Using positivity of $\omega$\ on $Z'$, this condition is sufficient to ensure that all values of the form $\omega \left( A [N, C_H] \right)$\ are now also small. Using~(\ref{eq:ApproxFactor}) we have 
\begin{eqnarray*}
\left| \omega \left( A [N, C_H] \right) \right| &\leq&  \sum_{n=0}^M \left| \omega \left( B_n\, {\rm ad}_{C_H}^{n+1} (N_0) \right) \right|
\\
&\leq&  \sum_{n=0}^M \sqrt{\omega \left( B_n^* B_n \right)} \sqrt{\omega \left( \left( {\rm ad}_{C_H}^n (N_0) \right)^* {\rm ad}_{C_H}^n (N_0) \right)}
\\
&\propto& \hbar \epsilon \sum_{n=0}^M \sqrt{\omega \left( B_n^* B_n \right)}\, \hbar^n   \propto \hbar \epsilon\ .
\end{eqnarray*}
More generally, following the same argument we get
\[
\left| \omega \left( A\, {\rm ad}_{C_H}^n (N)  \right) \right| \propto \hbar^n \epsilon \ .
\]

Furthermore, the ``smallness'' is locally approximately preserved along the gauge orbits generated by $C_H$\ in the sense that the derivatives of these conditions along the gauge flows are also of order $\epsilon$. We use~(\ref{eq:Flow}) to compute changes in values assigned by an almost positive $\omega$ along the gauge flow generated by $GC_H$\ for an arbitrary $G\in \mathcal{A}$
\begin{eqnarray*}
\left| \frac{d}{d \lambda} S_{GC_H} (\lambda)  \omega \left(  A [N, C_H] \right) \right|_{\lambda=0} &=& \frac{1}{\hbar} \left| \omega \left( \left[ A [N, C_H], B C_H \right] \right) \right|
\\
&=& \frac{1}{\hbar} \left| \omega \left( B A\,  {\rm ad}_{C_H}^2 (N) + B [A, C_H] [N, C_H] \right) \right|
\\
&\leq& \frac{1}{\hbar} \left| \omega \left( B A\,  {\rm ad}_{C_H}^2 (N) \right) \right|+ \left|\omega \left( B [A, C_H] [N, C_H] \right) \right|
\\
&\propto& \hbar \epsilon \ ,
\end{eqnarray*}
which is of the same order as $\omega \left(  A [N, C_H] \right)$.

Depending on the particular system studied, this approximation can be quite manageable. For example, if $N=a \mathbf{1} + b Z^2$, as we considered earlier, we have $N_0=N \in Z'$, as well as
\[
{\rm ad}_{C_H} (N_0) =2i\hbar b Z \ ,\ \  {\rm ad}_{C_H}^2 (N_0) =-2\hbar^2 b \mathbf{1}\ , \ \ {\rm and} \ \ {\rm ad}_{C_H}^n (N_0) = 0 \ , \ \ {\rm for} \ \ n>2 \ .
\]
So~(\ref{eq:ApproxFactCond}) results in only two conditions for approximate factorization
\[
|b| |\omega(Z)| \propto \epsilon  \quad {\rm and} \quad |b| \propto \epsilon .
\]
It is then sufficient to require $b$\ to be small and for $\omega(Z)$\ to not be very large.

\subsection{Deparameterization by linearization}

Given a constraint $C$\ and a clock $(Z, \mathcal{F})$\ such that the commutator $[Z, C] \neq 0$, but is not of the canonical form, the idea here is to perform some approximately reversible transformation on $C$\ that will make it exactly deparameterizable with respect to $Z$
\[
LC=C_H \ .
\]
Our starting assumption will be that $L$, just as the form of the above expression suggests, is a left multiplication by some combination of elements of the kinematical algebra $\mathcal{A}$. In the most straightforward situation $L\in \mathcal{A}$\ and linearization is the reverse of factorization discussed in section~\ref{sec:Factorization} with $L=N^{-1}$. In fact, this relation immediately highlights where linearization requires more subtlety than factorization: in all of our explicit examples so far, with kinematical algebra $\mathcal{A}$\ constructed out of polynomials in basic (usually canonical) generators, no elements of $\mathcal{A}$\ other than multiples of the identity are invertible within $\mathcal{A}$\ itself. In a Hilbert space representation of $\mathcal{A}$\ inverses of some operators can be constructed by spectral decomposition. When an inverse operator $A^{-1}$\ exists, when we restrict to the overlap of domains of $A^n$\ for all positive integer $n$\ and $A^{-1}$, its action  coincides with the action of the infinite power series constructed out of $A$ 
\[
A^{-1} = \left( a_0\mathbf{1} + \left(A-a_0\mathbf{1}\right) \right)^{-1} = \frac{1}{a_0}\sum_{n=0}^\infty \left(\frac{-1}{a_0}\right)^n \left(A-a_0\mathbf{1} \right)^n \ ,
\]
for an arbitrary number $a_0$. The first approximation that we will employ here is the use of such formal power series to invert elements of $\mathcal{A}$: the results of our manipulations will only be defined on a subset of algebraic states, for which computing $\omega(A^{-1})$\ and related expressions converges.

For our purposes, it is convenient to define state-dependent ``moment'' elements
\begin{equation}\label{eq:MomentOperator}
\Delta  A = A- \omega(A) \mathbf{1} \ .
\end{equation}
For any fixed state this gives an element of $\mathcal{A}$, which will be 
different if a different state is selected. We will keep this state-dependence in mind, while omitting explicit reference to the state when writing $\Delta  A$\ to reduce notational clutter. By setting $a_0=\omega(A)$\ we re-write the inverse power series in terms of the moments
\begin{equation}\label{eq:InverseSeries}
A^{-1} = \frac{1}{\omega(A)}\sum_{n=0}^\infty \left(\frac{-1}{\omega(A)}\right)^n (\Delta  A)^n \ .
\end{equation}
Linearization may also involve inverting other polynimial operations, such as taking square roots, which can also be constructed using power series in moments. We, therefore, expect that linearization will have the general form of multiplication by some
\[
L =\sum_{n=0}^\infty f_n (\omega) A_n \ ,
\]
where $f(\omega)$\ are some state-dependent numerical coefficients and $A_n \in \mathcal{A}$. We also anticipate that the resultant linearized constraint $C_H$\ will also be of this form.

\subsubsection{Time independent quadratic constraint}

To show how linerization can be constructed in practice we will focus on the situation where $Z$\ is canonical, as in section~\ref{sec:LinearConstraintAlgebra}, and the constraint has the special quadratic form
\[
C=E^2-h\ ,
\]
with $h^*=h$\ and $[E,h]=0$. This is a special case of the quadratic constraint in equation~(\ref{eq:CompleteSquare}) with $\tilde{E} = E$, so if $h$\ has a square root we can factorize $C=(E-\sqrt{h})(E+\sqrt{h})$. Looking for linearization $LC=E+\sqrt{h}=:C_H$\ we will need to take the square root of $h$\ and invert $(E-\sqrt{h})$. For the kinematical algebra constructed out of polynomials in $E$\ and other generators, $h$\ will not in general have a square root in $\mathcal{A}$\ and even if it does, $(E-\sqrt{h})$ does not have an inverse in $\mathcal{A}$. We therefore use state-dependent moments to write
\begin{equation}
 E-\sqrt{h} = \omega(E-\sqrt{h}) + \Delta (E - \sqrt{h}) \ .
\end{equation}
For the states on which $h$\ has a square root and $(E-\sqrt{h})$\ can be inverted, enforcing $C$\ is equivalent to enforcing $C_H$, so that for a solution of $C$\ we also have $\omega(E-\sqrt{h})=\omega(2E-(E+\sqrt{h}))=2\omega(E)$. We define $L$ as
a formal state-dependent power series expanding $(E-\sqrt{h})^{-1}$ around $\omega(E)$\ as in equation~(\ref{eq:InverseSeries}):
\begin{equation} \label{N}
 L:=\frac{1}{2\omega(E)} \sum_{n=0}^{\infty}
 \left(-\frac{\Delta E-\Delta \sqrt{h}}{2\omega(E)}\right)^n\,.
\end{equation}
By construction, $LC=E-\sqrt{h}$ is linear in $E$ and $*$-invariant, where if $h$\ does not have a square root in $\mathcal{A}$, it can be constructed through its own moment power series
\[
\sqrt{h} = \sqrt{\omega(h) + \Delta h} = \sqrt{\omega (h)} \sqrt{1+\frac{\Delta h}{\omega(h)}} = \sqrt{\omega (h)} \sum_{n=0}^\infty {\frac{1}{2} \choose n} \left( \frac{\Delta h}{\omega(h)}\right)^n \ .
\]

\subsubsection{Time dependent quadratic constraint}

Here the constraint is similar to the previous section $C=E^2-H^2$, but with $[E,H]\not=0$, so that we cannot factorize the constraint:
$E^2-H^2\not=(E-H)(E+H)$. However, we can find appropriate factors again at
the level of formal power series. To show this, we rewrite our constraint as
$C=E^2-H^2-V$ with $[E,H]=0$ but $[E,V]\not=0$, explicitly splitting off a
time-dependent potential $V$. In this form, there is the additional problem of
taking the square root of $H^2+V$. If the square root is not obtained from a
representation on a kinematical Hilbert space via the spectral decomposition
of $H^2+V$, it can be defined by a formal power series
\begin{equation} \label{sqrt}
 \sqrt{H^2+V}:= H+\frac{1}{2} H^{-1}V+\cdots
\end{equation}
if $H$ is invertible and commutes with $V$. 

Given a square root, we make an ansatz to factorize the constraint as
\begin{equation} \label{CXY}
 C=E^2-H^2-V= \left(E+\sqrt{H^2+V}+X\right)\left(E-\sqrt{H^2+V}+Y\right)
\end{equation}
with $X,Y\in {\cal A}$ to be determined so as to make the equation an
identity. Both $X$ and $Y$ should be $*$-invariant for the two terms in the
factorization (\ref{CXY}) to serve as either $N$ or $C_H$.

We compute
\begin{eqnarray*}
  C &=& E^2-H^2-V+\left[\sqrt{H^2+V},E\right]\\
  &&+ X\left(E-\sqrt{H^2+V}\right)+
 \left(E+\sqrt{H^2+V}\right)Y+XY \,.
\end{eqnarray*}
The terms in $C$ that are not manifestly $*$-invariant are given
by
\begin{equation}\label{CIm}
  0=C-C^*=\left[\sqrt{H^2+V},2E+X+Y\right]
  +[X-Y,E]+[X,Y]\,.
\end{equation}
The symmetric terms required to vanish for (\ref{CXY}) to be valid are
\begin{eqnarray} \label{CReal}
  0&=&(C+C^*)- 2(E^2-H^2-V)= \left((X+Y)E+E(Y+X)\right)\\ 
 && -\left((X-Y)\sqrt{H^2+V}+\sqrt{H^2+V}(X-Y)\right)+ XY+YX\,. \nonumber
\end{eqnarray}

It is difficult to find general solutions to these equations. One simple but
non particularly interesting special solution to (\ref{CReal}) is obtained if we set
$Y=-X$. Equation (\ref{CReal}) then implies
\[
  X^2+\left(X\sqrt{H^2+V}+\sqrt{H^2+V}X\right)=0
\]
with an obvious solution $X=-2\sqrt{H^2+V}$. This simple solution is of little
interest because it merely flips the two factors in (\ref{CXY}). Moreover,
because these two factors do not commute in the time-dependent case, it cannot
be a complete solution, and indeed equation (\ref{CIm}) is violated.

Nevertheless, assuming that $Y=-X$ is useful because it allows us to make
contact with previous work on effective constraints. At least formally, we can
factorize the constraint in the form (\ref{CXY}) if we do not insist on
$*$-invariant $X$ and $Y$. This condition is necessary in our algebraic
deparameterization because it guarantees a $*$-invariant flow rate $N$ and
Hamiltonian in $C_H$, such that $C=NC_H$. These invariance conditions, in
turn, are required for a well-defined flow that preserves the reality of
fashionables and is meaningful off-shell. If $X$ or $Y$ are no longer
$*$-invariant, at least one of these conditions must be violated.

If we then ignore the condition of $*$-invariance of $C$, the factorization
(\ref{CXY}) imposes only one equation that relates $X$ and $Y$, instead of two
equations, (\ref{CIm}) and (\ref{CReal}). We may again choose $Y:=-X$, such
that $C=E^2-H^2-V$ in (\ref{CXY}) implies the condition
\begin{equation}
 2\sqrt{H^2+V}X= \left[\sqrt{H^2+V},E\right]+ \left[X,E-\sqrt{H^2+V}\right]-X^2\,.
\end{equation}
We solve this equation for $X$, assuming $\sqrt{H^2+V}$ to be invertible and using
a formal power series,
\begin{eqnarray} \label{X}
 X&=&\frac{1}{2}\sqrt{H^2+V}^{-1} \left([\sqrt{H^2+V},E]+
   \frac{1}{2}\left[\sqrt{H^2+V}^{-1}[\sqrt{H^2+V},E],
     E\right]\right. \nonumber\\
&&\left.- \frac{1}{2} \sqrt{H^2+V}^{-1} [\sqrt{H^2+V},E]^2+\cdots\right)\,,
\end{eqnarray}
iteratively inserting $X$. The presence of iterated and squared commutators
means that this formal power series takes the form of an expansion by powers of
$\hbar$.

If $[V,H]=0$ (but still $[V,E]\not=0$), we can compute $[\sqrt{H^2+V},E]$ by
interpreting $\sqrt{H^2+V}$ as the formal power series (\ref{sqrt}). The
expansion in (\ref{X}) is then done by iterated commutators
$[\cdots[V,E],\cdots,E]$ which are derivatives of $V$ by $Z$. (It takes the
form of an adiabatic expansion.) For polynomial $V(Z)$, therefore, the formal
power series (\ref{X}) truncates after finitely many terms.
Using the assumption $[V,H]=0$, we have, to first order in $\hbar$ (or in commutators
with $E$),
\begin{equation}
 X=\frac{i\hbar V'(Z)}{4H^2}+O(\hbar^2)+O(V^2) 
\end{equation}
if the square root is understood as in (\ref{sqrt}). Clearly, this solution
for $X$ (and, correspondingly, for $-Y$) is not $*$-invariant.

The $C$-flow under the quantum clock $Z$, generated by
$LC=E-H-\frac{1}{2}H^{-1}V-X$, would be $*$-invariant if
$V(Z)+\frac{1}{2}i\hbar H^{-1}V'(Z)$ could be $*$-invariant. This condition
(for $H^*=H$) can be fulfilled only if $Z$ is {\em not} $*$-invariant. We
dismiss this possibility because $Z$ is a member of a basic canonical pair,
and its $*$-invariance is required for $Z'$ to inherit a $*$-structure from
$\mathcal{A}$. Since fashionables are defined as a subset of $Z'$, a
$*$-structure is required for a meaningful physical interpretation.

Alternatively, given the notion of almost-positive states, we may consider a
weaker condition on the flow generated by $NC$. Instead of requiring $LC$ to
be $*$-invariant, we can impose the condition that $\omega(LC)$ be real for
admissible states $\omega$. For an almost-positive state, $LC\not=(LC)^*$ does
not imply a non-real $\omega(LC)$ off-shell because $LC\not\in Z'$.  The
contribution in $LC$ that is not $*$-invariant is given by
$\frac{1}{2}H^{-1}V-X$ or, since we assume that $[V,H]=0$, by
$V(Z)+\frac{1}{2}i\hbar H^{-1}V'(Z)$.  If we then require that
$\omega(V(Z)+\frac{1}{2}i\hbar H^{-1}V'(Z))$ be real, $\omega(Z)$ cannot be
real. We may still assume that $Z$ is $*$-invariant if we modify our
definition of almost-positive states to be positive only on fashionables,
rather than the full $Z'$. A non-zero imaginary part of $\omega(Z)$ is
required by the $\hbar$-term in $X$, so that it should be of the order of
$\hbar$.

For a polynomial $V(Z)=\sum_nV_nZ^n$, we expand
\[
  \omega(V(Z))=\sum_n V_n \omega(Z^n)= \sum_nV_n\omega(Z)^n
\]
and
\[
  \omega(H^{-1}V'(Z))= \sum_n nV_n\omega(Z)^{n-1}\omega(H^{-1}) \,.
\]
Here, we use the fact that the condition $\omega(ZA)=\omega(Z)\omega(A)$ for
all $A\in\mathcal{A}$ if $\omega$ is almost-positive implies
$\omega(Z^n)=\omega(Z)^n$ because $Z\in\mathcal{A}$.  Writing
$\omega(Z)={\rm Re}\,\omega(Z)+i{\rm Im}\,\omega(Z)$ and treating
${\rm Im}\omega(Z)$ as a number of the order $\hbar$, we can solve for
\begin{equation} \label{Im}
 {\rm Im}\,\omega(Z)= -\frac{1}{2} \hbar \omega(H^{-1})+O(\hbar^2)\,.
\end{equation}
The imaginary part is thus fixed, and only the real part of $\omega(Z)$ plays
the role of an evolution parameter in the flow equation (\ref{eq:Hflow}). This
result agrees with what had been found previously using effective
constraints.

In this intrepretation, we may have $*$-invariant $C$, $L$, $C_H$ and $Z$. As
a consequence, off-shell gauge transformations and deparameterized evolution
are meaningful, and a natural $*$-structure is induced on $Z'$ which can be
applied to fashionables. However, if we use a complex $z$ instead of a real
$t$ for $\omega(Z)$, $(Z-z\mathbf{1})Z'$ does not define a $*$-invariant
subalgebra of $Z'$, and $Z'/(Z-z\mathbf{1})Z'$ does not inherit a natural
$*$-structure from $Z'$. The projection
$\pi_t\colon Z'\to Z'/(Z-z\mathbf{1})Z'$, used in the definition of the
gauge-fixing surface (\ref{gaugefixing}), is no longer a $*$-homomorphism. It
would therefore be imposible to define positivity conditions on gauge-fixed states.

The present derivation shows that we would have to violate at least one of the
conditions of algebraic deparameterization as defined here if we wanted to
implement previous results from effective constraints. It therefore remains
unclear whether complex time evolution with an imaginary contribution
(\ref{Im}) can be extended from semiclassical evolution to full quantum
evolution.

\subsection{Moments and effective constraints}

Effective constraints in canonical theories can be derived from moment
expansions, replacing functionals $\omega$ on an algebra with sets of
infinitely many moments. In semiclassical and perhaps other regimes, a finite
number of lower-order moments may be sufficient to describe the dynamics,
giving rise to systematic approximation methods.  This canonical version of
the method of effective actions, introduced and developed in
\cite{EffAc,EffCons,EffConsRel}, is rather close to the algebraic viewpoint of
the present article.

Given a basic set of algebra elements $x_i$ for $i=1,\ldots n$ which
generate the algebra ${\cal A}$ and have closed commutator relations with one
another, we define the moments
\begin{equation}
  \Delta (x_1^{a_1}\cdots x_n^{a_n}):=
  \omega\left((x_1-\omega(x_1))^{a_1}\cdots 
    (x_n-\omega(x_n))^{a_n}\right)_{\rm Weyl}
\end{equation}
of a given state $\omega$, indicating by the subscript ``Weyl'' that all
products of the $x_i$ are ordered completely symmetrically. (If the $x_i$ are
$*$-invariant, the moments are then real numbers in a positive state.)

\subsubsection{Uncertainty relations}

These moments are useful for different considerations of physical
properties. For instance, when one rewrites the Cauchy--Schwarz inequality in
terms of them, one obtains uncertainty relations. A standard derivation shows
that (\ref{CS}) for $A:=\Delta x_j=x_j-\omega(x_j)$ and
$B:=\Delta x_k$ implies
\begin{equation} \label{Uncertainty}
 \Delta (x_j^2)\Delta (x_k^2)- \Delta (x_jx_k)^2\geq
 \frac{|\omega([x_j,x_k])|^2}{4}
\end{equation}
if $\omega([x_j,x_k])$ is purely imaginary and $\Delta (x_jx_j)$ is
real. Otherwise, we would have
\begin{eqnarray*}
 \Delta (x_j^2)\Delta (x_k^2) &\geq& \left({\rm
                                                     Re}\Delta (x_jx_j)+
                                                     {\rm
                                                     Re}\omega([x_j,x_k])\right)^2\\ 
  &&+
 \frac{1}{4} \left({\rm
     Im}\Delta (x_jx_j)+ {\rm Im}\omega([x_j,x_k])\right)^2\,.
\end{eqnarray*}
For $A$ and $B$ polynomials in the $\Delta x_j$, we obtain
higher-order uncertainty relations for moments of order $a_1+\cdots+a_n> 2$
\cite{Casimir,ClassMoments}. 

It is interesting to note that moments of almost-positive linear functionals
formally satisfy the standard uncertainty relation even though they do not
have full positivity. The Cauchy--Schwarz inequality is replaced by
$\omega(ZA)=\omega(Z)\omega(A)$ for a quantum clock $Z$ if $A\in Z'$. From this
equation, we derive that $\Delta (Z^2)=0$, while
\begin{equation}
 \Delta (ZE)=\frac{1}{2}\omega(ZE+EZ)-\omega(Z)\omega(E)=
 \frac{1}{2}\omega([E,Z])
\end{equation}
is purely imaginary.  The uncertainty relation (\ref{Uncertainty}) therefore remains valid (and
saturated) for $x_j=Z$, $x_k=E$. (A vanishing fluctuation
$\Delta (Z^2)$ is formally consistent with the uncertainty relation
because $\Delta (ZE)$ is not real. This result is only formal because
the derivation of (\ref{Uncertainty}) is not valid if
$\Delta (x_jx_k)$ is not real, as it would be for $x_j=Z$ and
$x_k=E$.) This calculation confirms and explains analogous results derived for
effective constraints in an expansion to first order in $\hbar$
\cite{EffCons,EffConsRel}.

Semiclassical or $\hbar$-expansions are defined by the order of moments,
\begin{equation}
 \Delta (x_1^{a_1}\cdots x_n^{a_n})= O(\hbar^{(a_1+\cdots +a_n)/2})\,.
\end{equation}
(Since the commutator $[x_j,x_k]$ is proportional to $\hbar$, moments of
states that nearly saturate uncertainty relations must generically be of this
order.)  Any state with this behavior of the moments is called
semiclassical. For an algebra generated by a single basic canonical pair
$(q,p)$ with $[q,p]=i\hbar$, a special form of a semiclassical state is an
uncorrelated Gaussian state $\omega_{\sigma}$ for which
\begin{equation}
  \Delta _{\omega_{\sigma}}(q^ap^b)= 2^{-(a+b)}
  \hbar^a\sigma^{b-a}
  \frac{a!\,b!}{\left(a/2\right)!\left(b/2\right)!} 
\end{equation}
whenever $a$ and $b$ are even, and $\Delta _{\omega_{\sigma}}(q^ap^b)=0$
otherwise. (The fluctuation parameter $\sigma$ is of the order $\hbar/2$
because $\Delta (p^2)=\frac{1}{2}\sigma^2$.)

\subsubsection{Poisson structure}

The commutator in ${\cal A}$ induces a Poisson structure on the set of all
states by defining
\begin{equation}
 \{\omega(A),\omega(B)\}:= \frac{\omega([A,B])}{i\hbar}\,.
\end{equation}
It can be extended to polynomials in $\omega(x_i^k)$ by requiring the Leibniz
rule to hold, and then provides as Poisson structure on $\Gamma$, with
coordinates given by expectation values and moments of basic elements
$x_i$. An explicit calculation shows that
\begin{equation}
 \{\omega(x_i),\Delta (x_1^{a_1}\cdots x_n^{a_n})\}=0
\end{equation}
for canonical $x_i$. (There is a closed but lengthy expression also for
Poisson brackets of different moments \cite{HigherMoments}.)

On semiclassical states, one obtains a finite-dimensional Poisson manifold to
each order $N$, at which only moments with $a_1+\cdots+a_n\leq N$ are
considered; see \cite{Counting} for properties of such truncations. (These
manifolds, in general, are not symplectic.) By setting all higher-order
moments, as well as products of moments whose combined order exceeds $N$,
equal to zero, one obtains effective constraints
\begin{equation}
 C_{\rm p}:=\omega({\rm p}C)
\end{equation}
for all polynomials ${\rm p}$ in $x_i-\omega(x_i)$, of a certain maximum
degree required for moments up to order $M$. Poisson reduction of this
constrained system is equivalent to solving the constraint $C$ to order
$M/2$ in $\hbar$. The reduced phase space, given by the constraint surface
divided by the flow generated by the constraints, consists of observables of
the system to within the same order.

Alternatively, one may solve the system by fixing the gauge, that is, finding
a cross-section of the fibration given by the flow generated by effective
constraints. For effective constraints $C_{\rm p}$ with non-constant
${\rm p}$, a gauge-fixing condition is given by almost-positivity in the form
$\omega(ZA)=\omega(Z)\omega(A)$, which in terms of moments implies
\begin{equation}
 \Delta (Z^a\cdots)=0
\end{equation}
for all $a\geq 1$.
Up to constant multiples, only one effective constraint $\sum_{{\rm p}_i}
l^{{\rm p}_i}C_{{\rm p}_i}$ (expressed in a polynomial basis) remains unfixed,
which generates evolution
\begin{equation}
 \frac{{\rm d}\omega(F)}{{\rm d} z} = \{\omega(F),\sum_{{\rm p}_i} l^{{\rm
    p}_i}C_{{\rm p}_i}\}
\end{equation}
compatible with the gauge-fixing conditions:
\begin{equation}
 \{\Delta (Z^a\cdots),\sum_{{\rm p}_i} L^{{\rm
    p}_i}C_{{\rm p}_i}\}=0\,.
\end{equation}
This compatibility condition provides a set of linear equations for the
coefficients $l^{{\rm p}_i}$. (For examples, see
\cite{EffTime,EffTimeLong,EffTimeCosmo}.)  For several independent basic variables
$x_i$, or for higher orders in $\hbar$, these linear systems can become rather
large. 

Our general theory of algebraic deparameterization provides a more
practical method: We can expand $LC$ around $\omega(E)$, as in (\ref{N}). We
then recognize $\omega(LC)$ with (\ref{N}) as an expansion by effective
constraints. Coefficients in this expansion will therefore produce the
$l^{{\rm p}_i}$. To first order in $\hbar$, we include all effective
constraints with linear polynomials; $L$ in (\ref{N}) should therefore be
expanded to linear order. For a single canonical pair $(Q,P)$ in addition to
$(Z,E)$, we have
\begin{eqnarray}
 L &=& \frac{1}{2\omega(E)}
 \left(1-\frac{\Delta (E)+\Delta (H)}{2\omega(E)}+\cdots\right)\\
&=& \frac{1}{2\omega(E)}
 \left(1-\frac{\Delta (E)}{2\omega(E)}- \frac{1}{2H_{\omega}}
   \frac{\partial H_{\omega}}{\partial \omega(Q)} \Delta Q-
   \frac{1}{2H_{\omega}} 
   \frac{\partial H_{\omega}}{\partial \omega(P)} \Delta P
   +\cdots\right) \nonumber
\end{eqnarray}
where we introduced $H_{\omega}:= H_{\rm class}(\omega(Q),\omega(P))$ as the
classical limit of $H$, and Taylor-expanded in $\Delta Q$ and
$\Delta P$. Moreover, we identified $\omega(E)=H_{\omega}$ in
coefficients of $\Delta Q$ and $\Delta P$ using the
constraint, which is valid to this order of expansion. We then obtain the
linearized effective constraint
\begin{eqnarray}
 \omega(LC)&=&\sum l^{{\rm p}_i}C_{{\rm p}_i}\\
&=& \frac{1}{2\omega(E)} \left(C_1-
     \frac{1}{2\omega(E)} C_E- \frac{1}{2H_{\omega}}
   \frac{\partial H_{\omega}}{\partial \omega(Q)} C_Q- \frac{1}{2H_{\omega}} 
   \frac{\partial H_{\omega}}{\partial \omega(P)} C_P+\cdots \right)\nonumber
\end{eqnarray}
which generates a flow compatible with the quantum clock $Z$. This general expression
agrees with the specific examples found in
\cite{EffTime,EffTimeLong,EffTimeCosmo}. The new methods presented here offer a
streamlined derivation of effective evolution generators and, at the same
time, highlight a direct relationship between gauge-fixing conditions of
effective constraints and (almost-)positivity conditions on an algebraic state.

\section{Hilbert spaces}\label{sec:HSpace}

In the previous section, we have provided a well-defined scheme to obtain
complete physical evolution of algebraic states. We were able to avoid several
difficulties usually encountered when one attempts similar constructions on
physical Hilbert spaces. Nevertheless, it is often useful to have
Hilbert-space representations at hand. In our context, Hilbert spaces are
important in order to discuss convergence issues of the various formal power
series we referred to. These applications occur at two different levels,
amounting to kinematical and physical Hilbert spaces.

\subsection{Kinematical Hilbert space}

When one quantizes a theory in standard form, one constructs a $*$-algebra not
abstractly but rather as operators on a Hilbert space. Our algebra ${\cal A}$
would follow in this way by using a kinematical Hilbert space
representation. Classical observables (real functions on phase space) are
assigned self-adjoint operators, so that Poisson brackets of a set of basic
variables $x_i$ turn into commutators of their operators. One of the classical
phase-space functions is the constraint $C$, which becomes an element of
${\cal A}$ in the quantization procedure. So far, we have simply assumed that
some $*$-algebra with a set of basic element and a constraint is constructed
in this process, which we have analyzed further.

A kinematical Hilbert space provides different topologies in which one can
formulate the convergence of power series, such as (\ref{sqrt}). It also
allows one to analyze the invertibility of operators, such as $H$ in
(\ref{sqrt}), which would be more difficult at the pure algebra
level. Moreover, for normal operators, the spectral decomposition is a
powerful method to compute explicit versions of square roots or inverse
operators. All this can be done on a kinematical Hilbert space and does not
require one to enter the complicated issues surrounding the construction of
physical Hilbert spaces. We can therefore appeal to a kinematical Hilbert
space in the context of our formal power series. In particular, instead of
just assuming that a square root or an inverse exists in our algebra, we could
go back to the kinematical Hilbert space and enlarge our $*$-algebra by
elements corresponding to square roots or inverses of required operators.

The state-dependent expansion (\ref{N}) is less obvious to deal with. However,
in order to test convergence, we may simply view $\omega(E)$ in this series as
a (non-zero) number and discuss the convergence of the resulting operator
series. If the series converges to a well-defined operator for all non-zero
$\omega(E)$ (or at least some range), we can consider the limit operators as
elements of our $*$-algebra. Well-defined $N\in {\cal A}$ and thus evolution
generators $NC$ are then obtained.

The final kinematical question is about properties of the generator $NC$. By
our conditions, it must be self-adjoint. Two problems may arise in this
context: First, as we have seen algebraically, an $NC$ of the required form
can, in general, be self-adjoint only if $\omega(Z)$ is not real even though
$Z$ is self-adjoint on the kinematical Hilbert space. Secondly, even if the
series (\ref{N}) converges in a well-defined sense on the kinematical Hilbert
space, it may not do so uniquely. In the latter case, $NC$ would not be
essentially self-adjoint, even though it may have self-ajoint extensions.

The first problem has already appeared in the context of the state-dependent
expansion (\ref{N}). Heuristically, for physical observables we need only
require self-ajointness on the induced kinematical representation of the
sub-algebra of ${\cal A}$ ``not containing $Z$ and $E$.'' Using the conditions
on almost-positive states, we may then treat any occurrence of $Z$ in $NC$ as
a complex number $\omega(Z)$, and require that $NC$ be self-adjoint in this
representation for certain imaginary parts of $\omega(Z)$, such as
(\ref{Im}). However, as we have seen in our detailed algebraic discussion, a
full realization of the deparameterized system requires a meaningful choice of
a fashionable algebra which makes sense of the heuristic statement ``not
containing $Z$ and $E$'' above. Our definition of a fashionable algebra
requires a $*$-relation, but none would be inherited naturally if neither $Z'$
nor $Z/(Z-t\mathbf{1})Z'$ are $*$-invariant. This problem may not be obvious
on the Hilbert-space level, but is the underlying reason of several
uncontrolled ambiguities in this setting because the analog of a fashionable
algebra appears at best implicitly.

The second problem is not restricted to our formalism, and it can be dealt
with as usual. If an evolution generator is not essentially self-adjoint,
quantum dynamics is not considered uniquely determined. For all self-ajoint
extensions, we obtain well-defined (but mutually inequivalent) quantum
evolutions. For each of them one can use our formalism to make predictions,
and confront them with experiments. The self-adjoint extension might, in
general, depend on ${\rm Re}\,\omega(Z)$. Such an outcome would mean that there
could hardly be any predictivity because there would be free functions
undetermined by physical laws: the extension parameters are functions of
time. However, in concrete models the constraint would be given by a
differential operator, whose inequivalent self-adjoint extensions are
classified in terms of boundary conditions. If one uses the requirement that
the boundary condition itself should not depend on the time when it is posed,
allowed self-adjoint extensions do not depend on ${\rm Re}\omega(Z)$.

\subsection{Physical Hilbert space}

We do not need to refer to a physical Hilbert space in our
treatment. Nevertheless, we can construct one if we restrict our
almost-positive linear functionals $\omega_z$, solving evolution equations
within a given quantum clock $Z$, to fashionables, on which they become
positive. Applying the GNS construction for one such state then provides a
physical Hilbert space representation with a flow on it. However, there is no
natural unitary relation between such Hilbert spaces for different choices of
quantum clocks. Our algebraic treatment, by contrast, presents a unified
treatment and clearly reveals possible ambiguities, for instance in the choice
of fashionable algebras.

\section{Implications for the problem of time}\label{sec:Implications}

We have described the main features of a new algebraic theory of
deparameterization that highlights key mathematical properties underlying the
problem of time. Our constructions are based almost completely on the initial
kinematical $*$-algebra $\mathcal{A}$ used to define the system as well as
natural ingredients such as ideals, factor spaces, and homomorphisms. The only
exception to naturalness is the introduction of a fashionable algebra
$\mathcal{F}$ whose elements serve as observables evolving with respect to a
quantum clock $Z$. But even in this ingredient, there are well-defined
conditions that place it within the kinematical algebra.

Such a streamlined treatment of quantum symplectic reduction can be expected to
facilitate comparisons of physical results based on different choices of
internal times, possibly also of transformations between different times. In
the traditional treatment, by contrast, one would have independent physical
Hilbert spaces for different internal times, which in general are not related by any natural
transformations. A growing list of papers has shown physical
inequivalences between results obtained using different internal times, mainly
in cosmological systems
\cite{TwoTimes,MultChoice,BianchiInternal,SingClock,ClockDep}. This outcome is
not surprising if one considers the largely uncontrolled set of choices
that enter the definition of a physical Hilbert space in its usual
derivation. One may hope that a construction that clarifies and, as much as
possible, avoids such choices would make it easier to find transformations
between different versions and to make sure that their physical results
agree. Our algebraic definition of deparameterization is a first step in this direction (which at the moment runs in parallel with the already-mentioned representation-based approach of~\cite{QuantumRefSwitch,QuantumRef4}).

As it turns out, our new definition places strong conditions on any
well-defined implementation of evolution relative to an internal time, which had not been anticipated in the traditional
treatment of physical Hilbert spaces. Qualitatively, this observation confirms the
expectation that unified constructions in which different choices of quantum
clocks can consistently be compared should be more restricted than individual
quantizations based on a single clock, because transformability between
different clocks amounts to a physical invariance that is easily violated if
each instance is considered in isolation. Somewhat unfortunately, however, the
conditions appear to be very strong, making it hard to find a sufficiently
large number of interesting cosmological realizations.

All new conditions are a consequence of properties related to the
factorization $C=NC_H$ of the initial constraint operator $C$ in order to
obtain an evolution generator $[Z,C_H]=i\hbar \mathbf{1}$, which for a canonical $Z$\ corresponds to $C_H$\ being linear in its conjugate momentum. This factorization is crucial and appears, in some form, in
any derivation of a physical Hilbert space in which relational evolution is
formulated by a Schr\"odinger-type equation. It is perhaps not surprising that
our new strong conditions appear in relation to this factorization: 
\begin{itemize}
\item We require the constraint operator $C$ to be $*$-invariant, $C^*=C$ in
  order to have well-defined off-shell gauge flows that preserve the
  $*$-relation, with Dirac observables inheriting the $*$-structure from $\mathcal{A}$.
\item We require $C_H$ to be $*$-invariant, $C_H=C_H^*$, in order to have
  well-defined relational evolution that preserves the $*$-relation.
\item We assume constant flow rate, $[N,C_H]=0$, in order to be able to
  demonstrate that the values of observables are preserved by evolution
  generated by $C_H$, as detailed in Appendix~\ref{app:ConstantFlow
    Lemma}. This demonstration, in turn, is required because $C$ and $C_H$ do
  not have the same gauge orbits if $N$ is not invertible, a common feature in
  cosmological models in particular when the orignal constraint, $C$, is not
  linear in the time momentum $E$.
\end{itemize}
These conditions imply that the flow rate, $N$, must also be $*$-invariant
because $C=C^*$ with $C=NC_H$ where $C_H^*=C_H$ directly leads to $C_H
(N^*-N)=[N,C_H]$. Therefore, for constant flow rate we have $N^*=N$. All three
algebra elements in the equation $C=NC_H$ must therefore be $*$-invariant,
even while we have to keep $C_H$ to the right of $N$ in order for it to act
directly on a state.

A direct calculation attempting to factorize constraints quadratic in $E$
showed that $*$-invariance of $C$\ and its factors may be easier to implement if complex internal
times were allowed. Algebraically, we introduce internal time by a quantum
clock $Z\in\mathcal{A}$. Crucial steps of our construction are based on the
commutant $Z'\subset\mathcal{A}$ of $Z$, as well as the ideal
$(Z-t\mathbf{1})Z'\subset Z'$ and the factor space of the former by the
latter. We use these ideals and factor spaces in order to define evolving
observables (fashionables) and gauge-fixing surfaces that correspond to fixing
the internal time value $t$ of the quantum clock $Z$. For well-defined
observables and gauge fixings, we need $*$-operations on $Z'$ and
$Z'/Z-t\mathbf{1}$ with a number $t$, which can then also be used to inrtroduce
positivity conditions on physical states. For $Z'$ to inherit a natural
$*$-operation from $\mathcal{A}$, $Z$ must be $*$-invariant. Since $Z$ would
not be used to act on the physical Hilbert space, where it would rather be
replaced by the internal time $t$, physical states $\omega$ on $\mathcal{A}$
need not be positive or real when applied to $Z$. It may therefore be possible
to have a complex expectation value $\omega(Z)$ even if $Z$ is
$*$-invariant. If we use this complex value for our number $t$ that defines
internal time, however, $Z-t\mathbf{1}$ is not $*$-invariant and there is no
natural $*$-operation on the factor space $Z'/(Z-t\mathbf{1})Z'$. In algebraic
deparameterization, therefore, complex time is not a possibility to avoid
strong conditions on allowed models.

Most of these basic requirements cannot be circumvented because they are directly
related to $*$-operations, which in turn are needed for unambiguous definitions
of positivity of states (or, in more physical terms, uncertainty
relations). The only exception is the assumption of constant flow rate,
$[N,C_H]=0$, which we made for more technical reasons. If the relevant lemmas
can be generalized to non-constant flow rates, conditions on well-defined
deparameterizable constraints may be relaxable. 

Alternatively, because our new restrictions become especially confining in
relativistic systems with constraints non-linear in the time momentum $E$ and
with non-constant lapse functions $N$, they may be telling us something about
quantum time in relativistic gravitational systems. In general, when spacetime
geometry itself is subject to some form of quantum dynamics, one might expect
that the ability to interpret a model universe as a quantum system evolving
unitarily relative to an internal clock and even the existence of suitable
internal clocks may be limited. They may instead be approximate
emergent properties of a special class of dynamical solutions.

The semiclassical analysis of \cite{EffTime,EffTimeLong} has sometimes been
interpreted \cite{TimeShape} to mean that deparameterization \emph{requires}
the clock to be semiclassical or the quantum state of the universe to be
peaked around a classical trajectory (or ``history''). Our new algebraic
analysis of deparameterization provides a non-semiclassical vantage point on
this issue, with gauge-fixing conditions of effective constraints now
recognized as a general almost-positivity requirement on allowed states of the
quantum constrained system. Does a clock need to be semiclassical to
deparameterize a quantum constraint? This question is moot when asked within
deparameterization relative to a given clock $Z$, since within the reduction
it is just a parameter with vanishing spread. In the context of a physical
state that can be interpreted relative to multiple clocks $Z_1$\ and $Z_2$\
(for example if $C=E_1+E_2+H$, with $[H, Z_1]=[H, Z_2]=0$) there is nothing
precluding $Z_1$\ from being ``very quantum'', in the sense of having a large
spread or strong correlations with other variables, in states deparameterized
relative to $Z_2$\ and vice versa. The theory for transforming states from
$Z_1$ Zeitgeist to $Z_2$ Zeitgeist and back is yet to be worked out in the
algebraic approach, however there is no reason to believe that there will be
any restriction of states that can be transformed in the simplest cases of a
constraint exactly deparameterizable by two clocks. The situation becomes more
subtle if deparameterization is approximate and subject to localization
restrictions on states such as the ones discussed in
section~\ref{sec:ApproxFactorization}. In this scenario, the state still need
not be semiclassical, but does need to be localized to the region where the
given clock choice is valid.

\section*{Acknowledgements}

This work was supported in part by NSF grant PHY-2206591.

\appendix

\section{Result for factorization with a constant flow rate}
\label{app:ConstantFlow Lemma}

Let $C=NC_H$\ be deparameterized by factorization with respect to some clock $(Z, \mathcal{F})$\ as described in section~\ref{sec:Factorization}, such that the flow rate is constant $[N, C_H] =0$. Let $\omega$\ be almost positive with respect to the deparameterization of $C_H$\ by $Z$. Furthermore let $\omega$\ be such that left multiplication by $N$ within $Z'$\ can be canceled in $\omega$. We show that $\omega(O)$\ is constant along all flows generated by $\mathcal{A}C_H$\ for any physical observable $O \in \mathcal{A}_{\rm obs}$\ of the original constraint (that is, $O\in\mathcal{A}$\ and $[O, C]=0$).

First we show by induction that for any $O \in \mathcal{A}_{\rm obs}$\ and for all integers $n \geq1$ we have (using notation for repeated commutators introduced in section~\ref{sec:FactorObservables})
\begin{equation} \label{eq:LemmaInduction}
\left( {\rm ad}_{C_H}^{n-1} \left( [O,N] \right) \right) C_H + N {\rm ad}_{C_H}^{n} O = 0 \ .
\end{equation}
The case of $n=1$\ follows immediately from $[O, C] = [O, N C_H] = 0$. Assuming relation~(\ref{eq:LemmaInduction}) holds up to some $n$, we take a commutator with $C_H$\ on both sides of equation~(\ref{eq:LemmaInduction}) and use the fact that $[N, C_H]=0$\ to get 
\[
\left( {\rm ad}_{C_H}^n \left( [O,N] \right) \right) C_H + N {\rm ad}_{C_H}^{n+1} O = 0 \ ,
\]
thus completing the proof of~(\ref{eq:LemmaInduction}) by induction.

Now, using the algebra decomposition associated with deparameterization of $C_H$, an observable $O=O_1+O_2 C_H$, for some $O_1 \in Z'$\ and $O_2 \in \mathcal{A}$. So that ${\rm ad}_{C_H}^{n} O = {\rm ad}_{C_H}^{n} O_1 + \left( {\rm ad}_{C_H}^{n} O_2 \right)C_H$, where ${\rm ad}_{C_H}^{n} O_1 \in Z'$, since $O_1 \in Z'$\ as discussed in section~\ref{sec:LinearCGauge} (immediately following the definition of deparameterization). We re-write equation~(\ref{eq:LemmaInduction}) as
\[
\left( {\rm ad}_{C_H}^{n-1} \left( [O,N] \right) + {\rm ad}_{C_H}^{n} O_2 \right) C_H + N {\rm ad}_{C_H}^{n} O_1 = 0 \ ,
\]
where ${\rm ad}_{C_H}^{n} O_1 \in Z'$. It follows that for any $B\in Z'$ 
\[
\omega\left( B N {\rm ad}_{C_H}^{n} O_1 \right) = 0 \ , 
\]
where we have used the fact that $\omega$\ is almost positive. The left cancellation property of $N$\ within $Z'$\ now implies that for all $B\in Z'$\ we also have
\begin{equation} \label{eq:LemmaProperty1}
\omega \left( B \, {\rm ad}_{C_H}^{n} O_1 \right) = 0 \ .
\end{equation}

We now show that property~(\ref{eq:LemmaProperty1})
is preserved along an arbitrary constraint flow $S_{AC_H}(\lambda)$\ with any given $A\in \mathcal{A}$. We recall, that, since $Z'\cup \{C_H\}$\ algebraically generates $\mathcal{A}$\ can be uniquely written as a power series in $C_H$
\[
A= \sum_{m=0}^{M} G_m C_H^m \ ,
\]
for some integer $M$\ and some $G_m \in Z'$. According to lemma~8
of~\cite{AlgebraicTime}, the flow generated by $AC_H$\ on a solution to the
factor constraint $\omega\in \Gamma_{C_H}$ evaluated on any $F\in Z'$\
satisfies
\begin{equation}\label{eq:LemmaProjectedFlow}
i\hbar \frac{d}{d \lambda}\left(  S_{AC_H}(\lambda)\, \omega (F)\right) = S_{AC_H}(\lambda)\, \omega \left( \sum_{m=1}^{M+1} (-1)^{m-1} G_{m-1} {\rm ad}_{C_H}^m F \right) \ .
\end{equation}
Now we follow the logic of the proof of lemma~15 of~\cite{AlgebraicTime}: for every $B\in Z'$\ and $n\geq1$ let us define a function along the flow $f_{B}^{(n)} (\lambda) = S_{AC_H}(\lambda)\, \omega\left( B \, {\rm ad}_{C_H}^{n} O_1 \right)$. According to property~(\ref{eq:LemmaProperty1}), $f_{B}^{(n)} (0) = 0$, for all $B$\ and $n$. Now, suppose that all functions $f_{B}^{(n)} (\lambda') = 0$\ for some $\lambda'$, then, using equation~(\ref{eq:LemmaProjectedFlow}) we also have
\begin{eqnarray*}
i\hbar \left. \frac{df_{B}^{(n)}}{d \lambda}\right|_{\lambda=\lambda'} &=& \left. i\hbar \frac{d}{d \lambda}\left[  S_{AC_H}(\lambda)\, \left( B \, {\rm ad}_{C_H}^{n} O_1 \right)\right] \right|_{\lambda=\lambda'}
\\
&=& S_{AC_H}(\lambda')\, \omega \left[ \sum_{m=1}^{M+1} (-1)^{m-1} G_{m-1} {\rm ad}_{C_H}^m \left( B \, {\rm ad}_{C_H}^{n} O_1 \right) \right] 
\\
&=& \sum_{m=1}^{M+1} (-1)^{m-1} S_{AC_H}(\lambda')\, \omega \left[  G_{m-1}  \sum_{k=1}^{m} {m \choose l} \left( {\rm ad}_{C_H}^l B \right) \left( {\rm ad}_{C_H}^{n+m-l} O_1 \right) \right] 
\\
&=& \sum_{m=1}^{M+1} (-1)^{m-1} \sum_{k=1}^{m} {m \choose l}  f_{\left(G_{m-1}  \left({\rm ad}_{C_H}^l B\right) \right)}^{(n+m-l)} (\lambda')  =0 \ ,
\end{eqnarray*}
by our assumption, since every term in the sum is proportional to
$f_{B}^{(n)} (\lambda')$\ for some $B\in Z'$. Therefore
$\{ f_B^{(n)} (\lambda) = 0, \forall \lambda \}$\ is a solution to the flow
generated by $AC_H$\ on the functions we defined starting with the initial
state $\omega$. Since $A$\ is arbitrary
$S_{AC_H}(\lambda)\, \omega\left( B \, {\rm ad}_{C_H}^{n} O_1 \right) = 0$\
for all $A\in
\mathcal{A}$. 
Now the value of $\omega(O)$\ along the flow generated by $AC_H$\ for any $A\in \mathcal{A}$\ varies according to
\begin{eqnarray*}
i\hbar \frac{d}{d \lambda}\left(  S_{AC_H}(\lambda)\, \omega (O)\right) &=& i\hbar \frac{d}{d \lambda}\left(  S_{AC_H}(\lambda)\, \omega (O_1)\right)+ i\hbar \frac{d}{d \lambda}\left(  S_{AC_H}(\lambda)\, \omega (O_2 C_H)\right)
\\
&=& \sum_{m=1}^{M+1} (-1)^{m-1}  S_{AC_H}(\lambda)\, \omega \left( G_{m-1} {\rm ad}_{C_H}^m O_1 \right) 
\\
&&+\, S_{AC_H}(\lambda)\, \omega \left( [O_2, C_H] C_H \right) = 0 \ ,
\end{eqnarray*}
where we have used equation~(\ref{eq:LemmaProjectedFlow}) to evolve the value of $O_1$\ since it is in $Z'$, while we evolved the value of $O_2C_H$\ using the basic definition given in equation~\ref{eq:Flow}. The sum-over-$m$ term in the final expression vanishes by the immediately preceding argument, the second term vanishes because $\omega \in \Gamma_{C_H}$\ and therefore, according to lemma~4 of~\cite{AlgebraicTime}, $S_{AC_H}(\lambda)\, \omega \in \Gamma_{C_H}$. This completes the proof of invariance of the values assigned to $\mathcal{A}_{\rm obs}$\ by all states belonging to the orbit $[\omega]_{C_H}$.

\section{General form of a factorizable linear constraint}
\label{app:LinearC}

Consider a constraint $C$\ that is deparameterizable by factorization with respect to a clock $(Z, \mathcal{F})$, such that $[Z, [Z, C]]=0$. Therefore, there are $C_H, N \in\mathcal{A}$\, such that $C=NC_H$\ and $C_H$\ is a constraint that is directly deparameterizable by $Z$. 
To show that
\[
C=B_0 + B_1 C_H \ ,
\]
with $B_0, B_1 \in Z'$, we note that, by the definition of deparameterization
in section~\ref{sec:LinearCGauge}, $Z'\cup{C_H}$\ algebraically generate
$\mathcal{A}$, therefore there are some $M\in {\mathbb Z}$\ and $B_n\in Z'$\ (that is $[B_n,Z]=0$), such that
\[
C= \sum_{n=0}^M B_n C_H^n \ .
\]
Since $[Z, C_H]=i\hbar\mathbf{1}$, we have the usual result for conjugate variables
\[
[Z, C_H^n] = i\hbar n C_H^{n-1} \ .
\]
We then obtain
\begin{eqnarray*}
[Z, [Z, C]] &=& \left[ Z, i\hbar \sum_{n=1}^M n B_n C_H^{n-1} \right] 
\\
&=& -\hbar^2 \sum_{n=2}^M n (n-1) B_n C_H^{n-2}
\\
&=& -\hbar^2 \left[ 2B_2 + \left(\sum_{n=3}^M n (n-1) B_n C_H^{n-3} \right) C_H \right] = 0 \ .
\end{eqnarray*} 
Since $Z'\cap\mathcal{A}C_H=\{0\}$, the two terms in the square parentheses in the final expression must vanish separately, so that $B_2=0$\ and $\left(\sum_{n=3}^M n (n-1) B_n C_H^{n-3} \right) C_H=0$. However, since $C_H$\ is not a divisor of zero, this also gives
\[
\sum_{n=3}^M n (n-1) B_n C_H^{n-3} = 6B_3 + \left(\sum_{n=4}^M n (n-1) B_n C_H^{n-4} \right) C_H=0 \ .
\]
We repeat the above argument to conclude that $B_3=0$\ and $\sum_{n=4}^M n (n-1) B_n C_H^{n-4} =0$. Continuing to iterate we conclude that $B_n=0$\ for all $n\geq2$, obtaining $C=B_0 + B_1 C_H$, with $B_0, B_1 \in Z'$, as desired.

We now want to show that for this factorization $[N, Z] = 0 = [N, C_H]$\ and $N=N^*$. To do so, we set the factorized and power-series forms of the constraint equal $NC_H=B_0 + B_1 C_H$, so that
\[
B_0 + (B_1-N) C_H = 0 \ .
\]
Again, the two terms must vanish separately, so that $B_0=0$\ and $(B_1-N) C_H = 0$. Again, since $C_H$\ is not a divisor of zero, this gives $N=B_1 \in Z'$. Now we have
\[
[Z, C] = [Z, NC_H] = i\hbar N \ .
\]
Taking the star-involution of both sides with $Z^*=Z$\ and $C^*=C$, this immediately gives $N^*=N$. Equation~(\ref{eq:AdjRelation}) then implies $[N, C_H] = 0$. Therefore, if $C$\ is deparameterizable by factorization and $[Z, [Z, C]]=0$, the constraint must have the form $C=NC_H$, where $N=N^*\in Z'$\ and $[N, C_H]=0$, as claimed.

%\bibliographystyle{../preprint}
%\bibliography{../Bib/QuantGra}

\end{document}